\pdfoutput=1

\documentclass[11pt]{article}

\usepackage[margin=1in]{geometry}
\usepackage[english]{babel}
\usepackage[compact]{titlesec}
\usepackage[T1]{fontenc}

\usepackage{authblk} 

\usepackage[svgnames,xcdraw,table]{xcolor}
\definecolor{myBlue}{rgb}{0.1,0.1,0.8}
\definecolor{DarkGreen}{rgb}{0.1,0.5,0.1}
\usepackage{hyperref}
\hypersetup{
	colorlinks=true,
	linkcolor=red,
	urlcolor=DarkGreen,
	citecolor=blue
}
\usepackage[hyperpageref]{backref}
\renewcommand*{\backref}[1]{}
\renewcommand*{\backrefalt}[4]{%
    \ifcase #1 (Not cited.)%
    \or        (Cited on page~#2)%
    \else      (Cited on pages~#2)%
    \fi}

\usepackage{times}
\usepackage{soul}
\usepackage{url}
\usepackage{amsmath}
\usepackage{amsthm}
\usepackage{booktabs}

\usepackage{amsfonts}
\usepackage[inline]{enumitem}
\usepackage{etoolbox}
\usepackage{makecell}
\usepackage{mathtools}
\usepackage{multirow}
\usepackage[table]{xcolor} 
\usepackage{tabularx}
\usepackage{tabulary}
\newcolumntype{K}[1]{>{\centering\arraybackslash}p{#1}}
\usepackage{thmtools}
\usepackage{thm-restate}
\usepackage{dsfont}
\usepackage[textsize=footnotesize,color=red!25,bordercolor=red]{todonotes}
\usepackage{wasysym}
\usepackage[capitalise,noabbrev]{cleveref}
\Crefname{remark}{Remark}{Remarks}
\Crefname{rmk}{Remark}{Remarks}
\Crefname{dfn}{Definition}{Definitions}
\Crefname{thm}{Theorem}{Theorems}
\Crefname{cor}{Corollary}{Corollaries}
\Crefname{lem}{Lemma}{Lemmas}
\Crefname{examplex}{Example}{Examples}
\Crefname{prop}{Proposition}{Propositions}

\usepackage{tikz}
\usetikzlibrary{shapes,fit,calc}

\colorlet{mygray}{gray!40}

\declaretheoremstyle[
    bodyfont=\normalfont
]{normalstyle}

\declaretheorem[style=normalstyle,name=Definition]{dfn}

\newcommand{\bn}{\mathbb{N}}
\DeclareRobustCommand{\Chi}{{\mathpalette\irchi\relax}}
\newcommand{\irchi}[2]{\raisebox{\depth}{$#1\chi$}}

\newcommand{\mi}{\mathcal{I}}
\newcommand{\ns}{\text{ns}}

\usepackage{color}

\usepackage{natbib}

\usepackage{MnSymbol}

\newcommand{\EFtt}[1]{\ifstrempty{#1}{\texttt{\textup{EF}}}{\texttt{\textup{EF{#1}}}}}
\newcommand{\largeqtt}{\texttt{LARGE}}
\newcommand{\MMStt}{\texttt{MMS}}
\newcommand{\Proptt}[1]{\ifstrempty{#1}{\texttt{\textup{PROP}}}{\texttt{\textup{PROP{#1}}}}}
\newcommand{\smallqtt}{\texttt{SMALL}}

\newcommand{\EF}[1]{\ifstrempty{#1}{\textrm{\textup{EF}}}{\textrm{\textup{EF{$#1$}}}}}
\newcommand{\MMS}[1]{\ifstrempty{#1}{\textrm{\textup{MMS}}}{\textrm{\textup{MMS{$#1$}}}}}
\newcommand{\Prop}[1]{\ifstrempty{#1}{\textrm{\textup{PROP}}}{\textrm{\textup{PROP{$#1$}}}}}

\newcommand{\solo}{solo}
\newcommand{\Solo}{Solo}
\newcommand{\bundled}{full}
\newcommand{\Bundled}{Full}

\newcommand{\private}{private}

\newcommand{\public}{public}

\newcommand{\implicit}{implicit}

\newcommand{\explicit}{explicit}

\usepackage[capitalise,noabbrev]{cleveref}

\title{Bridging Theory and Perception in Fair Division: A Study on Comparative and Fair Share Notions}

\author[1]{Hadi Hosseini}
\author[2]{Joshua Kavner}
\author[1]{Samarth Khanna}
\author[3]{Sujoy Sikdar}
\author[4]{Lirong Xia}
\affil[1]{Pennsylvania State University}
\affil[2]{Rensselaer Polytechnic Institute}
\affil[3]{Binghamton University}
\affil[4]{Rutgers University}

\begin{document}

\maketitle

\begin{abstract}
The allocation of resources among multiple agents is a fundamental problem in both economics and computer science. In these settings, fairness plays a crucial role in ensuring social acceptability and practical implementation of resource allocation algorithms. Traditional fair division solutions have given rise to a variety of approximate fairness notions, often as a response to the challenges posed by non-existence or computational intractability of exact solutions. However, the inherent incompatibility among these notions raises a critical question: which concept of fairness is most suitable for practical applications? In this paper, we examine two broad frameworks---threshold-based and comparison-based fairness notions---and evaluate their perceived fairness through a comprehensive human subject study. Our findings uncover novel insights into the interplay between perception of fairness, theoretical guarantees, the role of externalities and subjective valuations, and underlying cognitive processes, shedding light on the theory and practice of fair division.
\end{abstract}



\section{Introduction}


\emph{Fair division} is the problem of allocating scarce resources to interested parties in a fair manner. 
Consider, for example, allocating different types of housing support to the homeless or refugees~\citep{abdulkadirouglu1998random}, computational resources to jobs in a computing cluster~\citep{ghodsi2011dominant}, or dividing an estate among heirs~\citep{Pratt90,brams1996fair}. 
In each application, the stakeholders want more of the resources, but to varying degrees. There are several fairness properties that could be sought after when determining the allocation, though it is unclear which principles should guide this process.


%

A rich theory of distributive justice offers compelling normative properties that formalize fairness in resource distribution problems. 
For instance, threshold-based properties guarantee that every agent receives a fair share and values their own bundle above a certain threshold. One example is {\em proportionality} (\Prop{}), which requires that each of $n$ agents receives a bundle worth at least $\frac{1}{n}$ their value for all goods~\citep{steinhaus1948problem}. Alternatively, comparison-based properties guarantee that each agent finds their bundle fair in comparison with others' bundles. One example is
{\em envy-freeness} (\EF{}), which requires that each agent prefers their bundle to that of any other agent~\citep{foley1966resource}.
%
%
However, this normative approach fails to address which properties \emph{ought} to be desired or which are representative of human values and decision-making. \citet{yaari1984dividing} argue that any satisfactory theory of distributive justice should be defensible against ``observed ethical judgments and moral intuitions,'' following Rawls's  \emph{reflective equilibrium} mode of analysis \citep{Rawls1971theory}. That is, moral principles should be represented by what people actually choose or find fair.

Economic experiments on distributional justice trace back at least to \citet{yaari1984dividing}, where human participants were offered resource allocations satisfying different fairness properties and asked which ones they found to be the most just. \citet{konow2003fairest} sought to evaluate fairness theories by how accurately they describe human fairness preferences. Herreiner and Puppe found that in bargaining games, human participants' choices of allocations are characterized more significantly by a combination of \emph{inequality aversion}--- i.e., seeking to reduce disparities between different agents and economic efficiency---than envy-freeness \citep{herreiner2007distributing,herreiner2009envy,herreiner2010inequality}.

Due to non-existence or computational challenges involved in finding \textit{exact} solutions (e.g. EF or PROP), the recent decade has seen a significant theoretical and algorithmic interest in designing approximate fairness notions, without much attention to their practical appeal.
In this line, recent works like~\citet{hosseini2022hide} have examined the perceived fairness of \textit{epistemic} approximations, which rely on information withholding, and \textit{counterfactual} approximations.
Their findings indicate that human participants generally were more satisfied with epistemic fairness notions compared to the counterfactual counterparts.
%


Our present work continues this line of experimental analysis about which distributive justice theories are representative of human decision-making.
We design and implement an experiment on Amazon’s Mechanical Turk platform to compare peoples’ perceptions of fairness of allocations satisfying various threshold- and envy-based fairness notions. This procedure enables us to address several research questions, such as which notions are perceived to be more fair, how sensitive perceived fairness is to contextual framing, and which allocation features contribute most to participant decision-making. Our research questions are described comprehensively, as follows.

\subsection{Research Hypotheses}

%

Proponents of threshold-based fairness guarantee conjecture that an allocation would be considered as fair if and only if agents receive a sufficient share of the resources. However, to the best of our knowledge, there is no empirically tested relationship between the the share of resources an agent receives and their perception of allocation fairness. This leads us to the following intuitive hypothesis.
%



\bigbreak
\noindent{\bf Hypothesis 1:} \emph{Increasing the fractional share of total resources received increases perceived fairness.}
\bigbreak

We test this hypothesis by offering participants varying shares of all resources, meeting different set thresholds, in their bundles and measuring perceived fairness \emph{explicitly} by asking whether they find their bundle acceptable or not. We find in Section \ref{sec:h1_perceived} that, perhaps unsurprisingly, perceived fairness is dependent and positively correlated with the amount of resources received. 

The challenge in allocating scarce and indivisible goods is that it is not always possible to guarantee that every agent receives a large share of their total value for all goods. Indeed, \Prop{} allocations do not always exist. \Prop{1} allocations, which approximate \Prop{}, always exists for any instance of the resource allocation problem with indivisible goods and additive valuations, i.e., when the value for a bundle is the sum of the values for goods in the bundle \citep{conitzer2017fair}. In a \Prop{1} allocation, the difference between an agent's proportional share and the value for their own bundle is at most the value of one good outside the agent's bundle. 

Our experiments show that, while the acceptability rate of \Prop{1} allocations is lower at $78\%$ than that of \Prop{} at $85\%$, this difference is {\em not} statistically significant.

\begin{figure}
    \centering
    \includegraphics[width=0.5\linewidth]{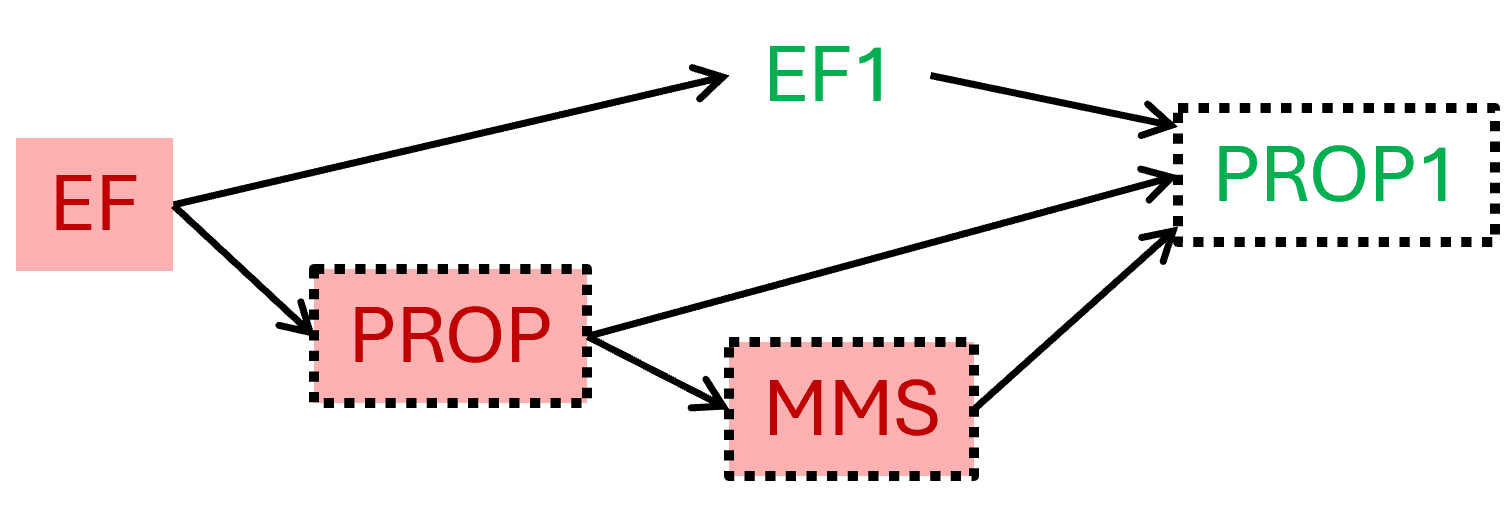}
    \caption{
    Relationships between different threshold- and envy-based fairness properties under non-negative and additive valuations. An arrow from property $X$ to property $Y$ represents the fact that if an allocation satisfies $X$, it must also satisfy $Y$. Threshold-based notions are outlined. Properties represented by red-shaded boxes may not exist for all fair division instances; allocations satisfying properties that are not shaded always exist.
    }
    \label{fig:relationships}
\end{figure}

For additive valuations, \EF{}, which guarantees no envy, also implies \Prop{}. Since \EF{} and \Prop{} allocations may not exist, the need to resort to their approximations results in the conundrum illustrated in \Cref{fig:relationships}, which summarizes the relationship between the different fairness measures we consider in this work. \Prop{}, and its relaxation \MMS{}, are incomparable with \EF{1}, which approximates \EF{}. This motivates our next hypothesis on the relative perceptions of the fairness of different comparative and threshold-based fairness properties.

\bigbreak
\noindent{\bf Hypothesis 2:} \emph{Comparative properties are perceived as more fair than threshold-based properties.}
\bigbreak

In Section \ref{sec:h2_threshold_vs_envy}, we find that there is not enough evidence to distinguish perceived fairness between threshold- and envy-based notions categorically. We find definitive evidence that \EF{} allocations are perceived as more fair than \EF{1} allocations. Surprisingly, we do not find sufficient evidence to establish a similar hierarchy over the threshold-based properties despite the theoretical hierarchy summarized in \Cref{fig:relationships}. Perhaps equally surprising is that \EF{1} and \Prop{1} are also indistinguishable despite \Prop{1} being strictly weaker than \EF{1}. We observe mixed results in distinguishing between \Prop{}, \MMS{} and \EF{1}.


Next, we explore how contextual attributes such as \textit{extrinsic information} and \textit{fairness measure}---which could be interpreted as \textit{framing} effects \citep{tversky1985framing}---influence  participants' responses.
In particular, we study the effects of providing the participant different amounts of \emph{bundle} and \emph{value information} by varying whether the participant is only  presented with information about their own bundle and valuations respectively, or is also aware of the bundles and valuations of other agents in the experiment.
In addition, we consider how \textit{explicit} and \textit{implicit} questions shape how fairness is measured. 
Inspired by threshold-based fairness notions, we ask {\em explicitly} whether the participant finds their bundle acceptable. 
In the spirit of comparative, envy-based notions, we provide participants the option of swapping their bundle with that of another agent, and the response provides an {\em implicit} measure of whether the participant found their bundle to be fair~\citep{hosseini2022hide}. This entails the following hypothesis.

\bigbreak
\noindent{\bf Hypothesis 3:} \emph{Perceived fairness is independent of extrinsic information about other agents and framing of the measure of fairness.
}
\bigbreak

We find in Section \ref{sec:h3_sensitivity_analysis} that participants perceive their bundle as fairer as (a) bundle information is increased, (b) value information is made private, and (c) fairness is measured explicitly rather than implicitly.
This suggests that perceived fairness indeed depends on information about other agents and experimental framing, confirming prior evidence that perceptions of algorithmic decisions are context-dependent \citep{lee18:understanding}.

Surprisingly, we also find that \EF{} is only perceived to be fairer than \Prop{} and \MMS{} at a statistically significant level by the implicit fairness measure, i.e., when we ask participants whether they wish to swap their bundle with that of another agent. Otherwise, when asked whether they find their bundles acceptable, there is insufficient evidence to distinguish between these properties despite \EF{} being the strictly stronger notion. Similarly, \Prop{} allocations are only perceived as more fair than \EF{1} allocations when valuations are private, while \MMS{} and \EF{1} remain indistinguishable.

Overall, we find that \EF{} allocations are perceived to be fair at a higher rate than allocations satisfying any of the fairness properties we consider, and the differences in rates of perceived fairness are statistically significant. While \EF{} allocations may fail to exist, as do \Prop{} and \MMS{}, our empirical approach and experimental results provide the practitioner with some useful insights. 

\begin{itemize}[leftmargin=*]
    \item Differences in the rates at which \EF{1} and \Prop{1} allocations are perceived to be fair compared to stronger notions like \EF{}, \Prop{} and \MMS{} are {\em not statistically significant}. Moreover, \EF{1} and \Prop{1} allocations always exist and can be computed in polynomial time when valuations are additive. 
    \item Our sensitivity analysis on the effects of information and framing shows that perceived fairness increases when allocations of all agents are revealed while agents' subjective valuations are kept private. Private valuations are a common feature of several practical resource allocation problems, and our results provide practitioners with guidance to make informed and application-specific decisions on the design of information in systems for performing resource allocation.
\end{itemize}





\vspace{0.5em}
\noindent{\bf Explaining perceived fairness reports and motivating swaps.} Building on these findings, we analyze the effect of various factors, beyond the theoretical fairness guarantees that  allocations provide, that potentially influence perceptions of fairness.

First, we consider the role that certain features of the allocation may have on participants' decision-making, including the fraction of the total possible value that participants receive and the amount of envy the participants experience.
Second, we categorize the swaps made by participants based on the effect of the swap on the distribution of welfare across all agents. For example, swaps may result in a Pareto improvement, benefiting both the participant and the other agent involved in the swap, or selfish, where the swap benefits the participant at the cost of another agent. We then determine the effect that value information has on participant decision-making for each of these categories. 

Our observations suggest that \textit{selfishness} is the primary motivation behind decisions to swap. Encouragingly, the frequency with which participants exercise the option to execute selfish swaps (when they are available) goes down when participants have more information about other agents' subjective valuations.

\vspace{0.5em}
\noindent{\bf Effect of information design and fairness measure on cognitive effort.} Finally, 
we investigate how 
our treatments of bundle information, value information, and fairness measure affect the cognitive effort exerted by participants in completing our questionnaire. It is known that cognitive effort is costly and humans often conserve cognitive effort to improve their subjective well-being \citep{westbrook2013subjective}. However, it is not clear whether humans will exert significant effort to verify fair outcomes, as studied in public goods games by \citet{fehr2002altruistic}. We therefore study whether cognitive effort, as measured by scenario completion time and reported difficulty, is dependent on treatment or correlates with perceived fairness. This is discussed further in Section \ref{sec:cognitive}.


\subsection{Related Work}

\vspace{0.5em}
\noindent{\bf Evaluating Theories of Fairness.} 
A large body of work examines how different allocation principles explain distributive preferences in a variety of settings \citep{konow2003fairest}. \citet{fehr1999theory} and \citet{bolton2000erc} provide different definitions of \textit{inequality aversion} and demonstrate their ability to explain behavior in ultimatum and dictator games. Similarly, Rawls' notion of \textit{maximizing the minimum} payoff \citep{Rawls1971theory}, along with a consideration for \textit{economic efficiency}, influences the choices of respondents in income-distribution scenarios when asked to choose among multiple options that determine the payoff for each recipient \citep{frohlich1987choices,charness2002understanding,engelmann2004inequality,cetre2019preferences}. When information about the identity of recipients is provided, the \textit{needs} of recipients appears to inform decisions in some scenarios \citep{gaertner2009primer} while the \textit{merit} or \textit{desert} of individuals plays a role in others \citep{Hoffman1985entitlements,overlaet1991criteria}. Additionally, it is observed that the relative preference over fairness notions depends on the cultural and demographic context \citep{murphy84factors}.

\vspace{0.5em}
\noindent{\bf Allocating Subjectively Valued Resources.}
\citet{yaari1984dividing} demonstrate how respondents' preferences over outcomes are influenced by the magnitude (intensity) and nature (needs, tastes, or beliefs) of the subjective utility recipients have over different resources.
\citet{herreiner2007distributing} observe that human subjects' preferences over allocations satisfying fairness properties such as EF, inequality aversion, and Rawlsian maximin, vary across different hypothetical instances where goods (and money) are to be distributed by a decision-maker. However, when agents are asked to arrive at mutually acceptable allocations through bargaining, the consideration for EF appears to be secondary to that for inequality aversion and Pareto optimality \citep{herreiner2010inequality}. When asked to choose out of a set of options for a shared item over which recipients have subjective preferences, respondents consistently select as per the maximin criterion, even across multiple rounds of decisions \citep{gates2020helpful}. In scenarios of rent division, where participants are assigned rooms and amounts of rent based on preferences over rooms, \citet{gal2016rent} show that EF allocations that also satisfy the maximin criterion are found significantly fairer than those that don't. 

\vspace{0.5em}
\noindent{\bf Fair Procedures Vs. Fair Outcomes.} A parallel line of work focuses on \textit{procedural justice}, or the study of fair procedures or \textit{methods}. People appear to choose more sophisticated allocation procedures such as the \textit{adjusted-winner}  \citep{brams1996fair} and \textit{Selfridge-Conway} \citep{robertson1998cake} mechanisms as compared to simpler procedures such as \textit{divide-and-choose} when provided with descriptions of each procedure \citep{schneider2004limitations,kyropoulou2022cake}. However, \citet{dupuis2009empirical,dupuis2011simpler} find that the allocations computed using more complex procedures are not perceived to be fairer, or more satisfying, than those computed by simpler methods such as the \textit{round robin} mechanism or alternative methods such as a \textit{genetic} algorithm. Multiple studies also observe that desirable properties, such as \Prop{} and \EF{}, which are otherwise guaranteed by the procedures being used, are lost due to strategic behavior from participants \citep{schneider2004limitations,daniel05bargaining,kyropoulou2022cake}. 

\vspace{0.5em}
\noindent{\bf Influence of Human Factors.} Perceptions of fairness depend on factors beyond the procedures used and outcomes achieved. \citet{Lee17:Mediation} observe that respondents find allocations decided through discussion with other participants fairer than those computed algorithmically on the Spliddit
\citep{goldman2015spliddit} platform for dividing items such as rent, tasks, and goods. Similarly, managerial decisions that require human skills, such as hiring and work evaluation, were found less fair if made by algorithms instead of humans, while this difference was not observed in mechanical decisions such as work assignment \citep{lee18:understanding}. \citet{Lee19:Justice} also demonstrate that allowing participants to \textit{change} an allocation provided by a platform like Spliddit significantly enhances perceived fairness while explaining the procedures and outcomes does not. \citet{uhde20shift} show that respondents prefer different distributive principles when asked about scenarios at an abstract level versus when given concrete examples. 

\section{Model and Solution Concepts}
\label{sec:model_defs}

\vspace{0.5em}
\noindent{\bf Model.}
For any $k\in \bn$, we define $[k]:=\{1,\dots,k\}$. An instance of the fair division problem is a tuple $\mi=\langle N, M, V \rangle$, where $N\coloneqq [n]$ is a set of $n$ {\em agents}, $M\coloneqq [m]$ is a set of $m$ {\em goods}, and $V\coloneqq \{v_1,\dots,v_n\}$ is a {\em valuation profile} that specifies for each agent $i\in N$ her preferences over the set of all possible {\em bundles} $2^M$. This {\em valuation function} $v_i:2^M\to\bn\cup\{0\}$ maps each bundle to a non-negative integer. 
We write $v_{i,j}$ instead of $v_{i}(\{j\})$ for a single good $j\in M$. 
We assume that valuation functions are {\em additive} so that for any $i\in N$ and $S\subseteq M$, $v_i(S)\coloneqq\sum_{j\in S}v_{i,j}$, where $v_i(\emptyset)=0$.
%
An \emph{allocation} $A\coloneqq(A_1,\dots,A_n)$ is a complete $n$-partition of the set of goods $M$, where $A_i\subseteq M$ is the \emph{bundle} allocated to agent $i\in N$. 

\begin{dfn}[\EF{}]
An allocation $A$ satisfies {\em envy-freeness} (\EF{}) if for every pair of agents $h,i\in N$, $v_i(A_i)\ge v_i(A_h)$~\citep{foley1966resource}.
\label{dfn:envy-free}
\end{dfn}

\begin{dfn}[\EF{1}]
An allocation $A$ satisfies {\em envy-freeness up to one good} (\EF{1}) if for each pair of agents $h,i\in N$, there exists a good $g_h\in A_h$ such that $v_i(A_i)\ge v_i(A_h \backslash \{g_h\})$~\citep{Lipton04:Approximately,budish2011combinatorial}.
\label{dfn:EF1}
\end{dfn}

\begin{figure}[t]
    \centering
    \includegraphics[width=0.65\textwidth]{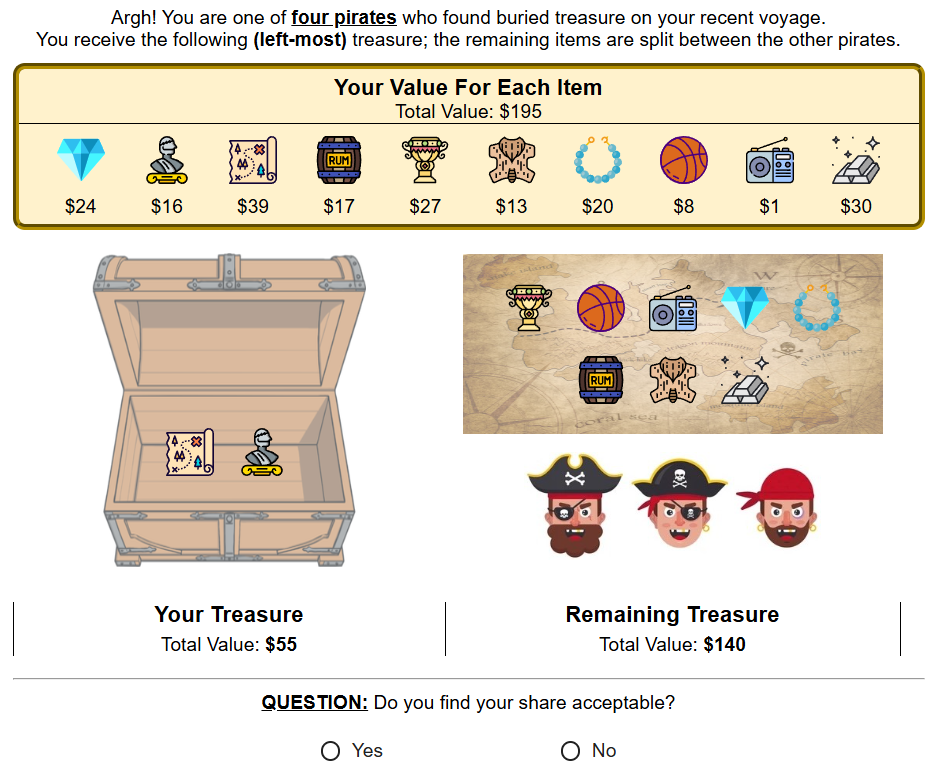}
    \caption{ 
        An example scenario of treatment T1, which provides \texttt{\private{}} information about valuations, \texttt{\solo{}} information about the allocation, and \texttt{\explicit{}} fairness measure. On top, the participant sees their value for each good and their total value for all goods. In the bottom left, the participant is presented with their own bundle as a collection of goods in their own pirate's chest, together with their total value for their bundle. Notice that the participant does not have any information about other agents' valuations, or how the remaining goods are allocated. Information in the bottom right about the participant's total additive value for all remaining goods is provided to facilitate comparisons with the value of the participant's bundle. Given this information, the participant is asked whether they find their share acceptable.
    }
\label{fig:sample_scenario_T1}
\end{figure} 

\begin{figure}[t]
\centering
\begin{tabular}{cc}
    \includegraphics[width=0.47\textwidth]{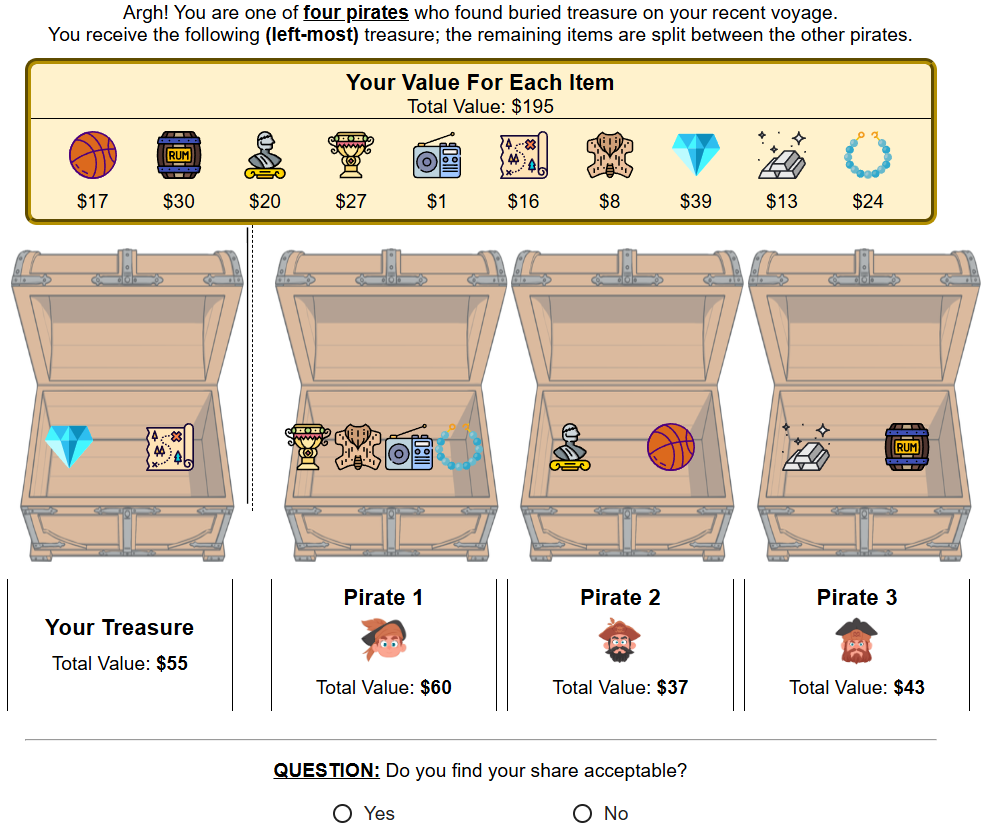}
     & 
    \includegraphics[width=0.47\textwidth]{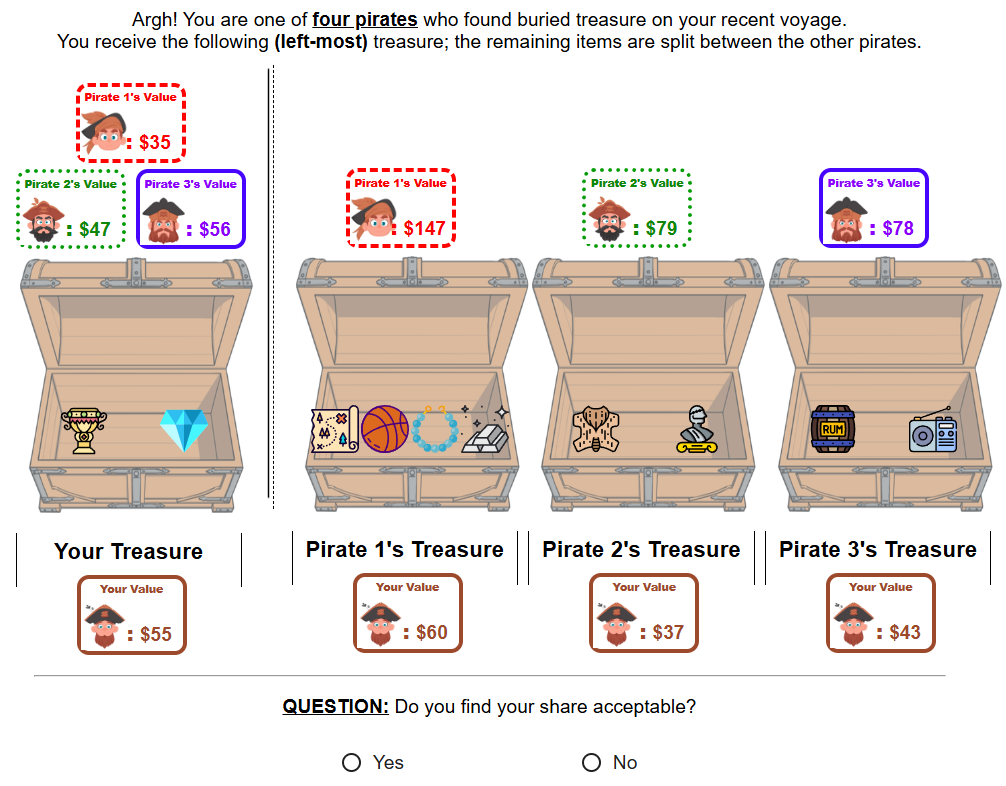}
     \\
    (a) & (b)
\end{tabular}
\caption{ 
    Examples of scenarios for treatments (a) T2 and (b) T4, illustrating how additional bundle and valuation information is presented. Treatment T2 (a) provides the participant with \texttt{\bundled{}} bundle information but \texttt{\private{}} valuation information. Information at the bottom allows the participant to compare their value for their own bundle with the bundle of every other agent. Treatment T4 (b) shows \texttt{\bundled{}} and \texttt{\public{}} information. Valuation information is made public by showing the participant how each other agent values both the their own and the participant's bundle immediately above the respective bundles. This facilitates judgments about how the participants' bundle may be perceived by other agents, as well as how swapping the participant's bundle with another agent will affect both parties. T3 and T5 are similar to T2 and T4, respectively, but with the implicit fairness measure instead.
}
\label{fig:sample_scenario_T2_T4}
\end{figure}

\begin{dfn}[\Prop{}]
An allocation $A$ satisfies
{\em proportionality} (\Prop{}) if $v_i(A_i) \geq \frac{1}{n} v_i(M)$ for every agent $i \in N$ \citep{steinhaus1948problem}.
\label{dfn:prop}
\end{dfn}

\begin{dfn}[\Prop{1}]
An allocation $A$ satisfies 
{\em proportionality up to one good} (\Prop{1}) if it is either \Prop{} or $\forall i \in N,~\exists h \in M \backslash A_i$ such that $v_i(A_i \cup \{h\}) \geq \frac{1}{n} v_i(M)$ \citep{conitzer2017fair}.
\label{dfn:prop1}
\end{dfn}

\begin{dfn}[\MMS{}]
For a set $S \subseteq M$, let $\Pi_n(S)$ be the set of $n$-partitions of $S$. The \emph{$n$-Maximin share guarantee} of agent $i \in N$ is
\[
MMS_i^{(n)}(S) = \max_{(T_1, \ldots, T_n) \in \Pi_n(S)} \min_{j \in N} v_i(T_j).
\]
Allocation $A$ is MMS if for all $i \in N$, $v_i(A_i) \geq MMS_i^{(n)}(M)$ \citep{budish2011combinatorial}.
\label{dfn:maximin}
\end{dfn}

\section{Experimental Design}

We designed an empirical study on Amazon Mechanical Turk to compare participants' perceived fairness of allocations satisfying one of seven threshold- or envy-based notions of fairness.

Participants were split into five information treatments and given ten scenarios in which they took the role of an agent in a fair division problem. Each scenario was gamified to represent pirates sharing buried treasure, such as jewelry or pelts (see e.g., Figure \ref{fig:sample_scenario_T1}). Participants were offered some information about each agent's bundle, as well as their own and others' value for the resources, and were then asked about their perceived fairness of their own bundle in one of two ways. 

Each treatment is characterized across three dimensions: bundle information, value information, and fairness measure, as we summarize in \Cref{tab:treatments}. {\em Bundle information} characterizes whether the participant is only aware of their own bundle (\texttt{\solo{}}) or knows every agent's bundle precisely (\texttt{\bundled{}}). \emph{Value information} characterizes whether the participant is only aware of their own values for the resources (\texttt{private}) or has some information about others' values too (\texttt{public}). \emph{Fairness measure} determines whether we measure perceived fairness by asking participants if they found their bundle acceptable (\texttt{explicit}) or asking which, if any, other pirate's bundle they wanted to swap with (\texttt{implicit}).

\begin{table}[ht]
    \centering
    \caption{The treatments participants in our study are subjected to are characterized by the amount of information about bundles and valuations that is revealed to the participant, and the framing of the question used to measure fairness. Bundle information is either \texttt{\solo{}}, meaning the participant only sees their own bundle, or \texttt{\bundled{}}, meaning all agents' bundles are visible. Valuations are either \texttt{\private{}}, or \texttt{\public{}} meaning the participant can determine how much each other agent values their own bundle and the participant's bundle. The question posed to the participant is either \texttt{\implicit{}}, measuring fairness by asking whether the participant would prefer to swap with another agent, or \texttt{\explicit{}}, asking whether the participant finds their bundle acceptable.}
    \begin{tabular}{|c||K{4em}|K{4em}|K{4em}|K{4em}|K{4em}|K{4em}|}  
    \hline
        \multirow{ 2}{*}{Treatment} &  \multicolumn{2}{K{8em}|}{Bundle Information} & \multicolumn{2}{K{8em}|}{Value Information} & \multicolumn{2}{K{8em}|}{Fairness Measure} \\ \cline{2-7}
         & \Solo{} & \Bundled{} & Private & Public & Implicit & Explicit \\ \hline
        T1 & $\checkmark$ & & $\checkmark$ & & & $\checkmark$\\ \hline
        T2 &  & $\checkmark$ & $\checkmark$ & & & $\checkmark$\\ \hline
        T3 & & $\checkmark$ & $\checkmark$ & & $\checkmark$ & \\ \hline
        T4 & & $\checkmark$ & & $\checkmark$ & & $\checkmark$\\ \hline
        T5 & & $\checkmark$ & & $\checkmark$ & $\checkmark$ & \\ \hline
    \end{tabular}
    \label{tab:treatments}
\end{table}

\vspace{0.5em}
\noindent{\bf Bundle information.}
With \texttt{\solo{}} information, participants were only informed about their own bundle of goods, whereas the remaining goods were distributed among the other agents (see e.g., Figure \ref{fig:sample_scenario_T1}(a)). This treatment enabled us to determine the relationship between the amount of resources the participant receives and perceived fairness, independent of what the other agents receive.
%
In contrast, with \texttt{\bundled{}} information, participants were informed of the complete allocation (see e.g., Figures \ref{fig:sample_scenario_T2_T4}(a) and (b)).

\vspace{0.5em}
\noindent{\bf Value information.}
Participants with \texttt{private}  information were only aware of their own \emph{subjective} additive values for the goods (see e.g., Figures \ref{fig:sample_scenario_T1} and \ref{fig:sample_scenario_T2_T4}(a)). They were taught in a tutorial (described in Section \ref{sec:survey_outline}, below) that their value is ``the sum of values of the items in the treasure.'' In contrast, participants with \texttt{public} information were additionally informed how each other agent values their own bundle and the participant's bundle (see e.g., Figure \ref{fig:sample_scenario_T2_T4}(b)). They were taught that their value ``may differ from how any other pirate values the treasure.''

\vspace{0.5em}
\noindent{\bf Fairness measure.}
Asking participants whether they perceive their bundle as ``fair,'' directly, is ill-defined and may be understood differently across participants. On the other hand, only using a single measure of perceived fairness suggests our results may not be broadly applicable. We therefore utilized two measures of perceived fairness. First, 
under \texttt{explicit} treatments, we asked
participants whether they found their share acceptable. This measure is inspired by the threshold-based definition of fairness, which assumes agents find their bundle fair if its value is large enough. 
Second, under \texttt{implicit} treatments, we asked participants if they wanted to keep their assigned treasure or to identify another agent's bundle to swap with. A participant's decision to swap suggests that they envy another agent's bundle. Under the envy-based definition of fairness, this reflects that they do not find their bundle fair.
In our analysis, we use ``yes'' and ``no swap'' responses to indicate a participant's perceived fairness of their allocation according to these respective measures.

\vspace{0.5em}
\noindent{\bf Survey details.}
Our study employed 30 mutually exclusive participants for each of five Human Intelligence Tasks (HITs), corresponding to the five treatments, in Amazon’s Mechanical Turk platform, totaling 150 participants. Our study was single-blind; participants were not aware of their treatment. Participants were paid a fixed amount, $\$1.00$, for completing the questionnaire, independent of their choices. This design choice is discussed in Section \ref{sec:limitations}.

\subsection{Data Set}
\label{sec:data_sets}

We developed a novel data set of $20$ instances and $7$ allocations per instance for a total of $140$ scenarios. 
%
%
%
Instances consisted of $4$ agents and $10$ goods. 
Valuations $v_{i,j}$ for each agent $i \in N$ and good $j \in M$ were sampled to be positive integers uniformly at random from $\{1,2,\ldots,50\}$.
Allocations were generated for each instance with rejection sampling by distributing the goods uniformly at random  until an allocation satisfying the desired fairness property was obtained.

\begin{figure}[t]
    \centering
    \includegraphics[width=0.85\linewidth]{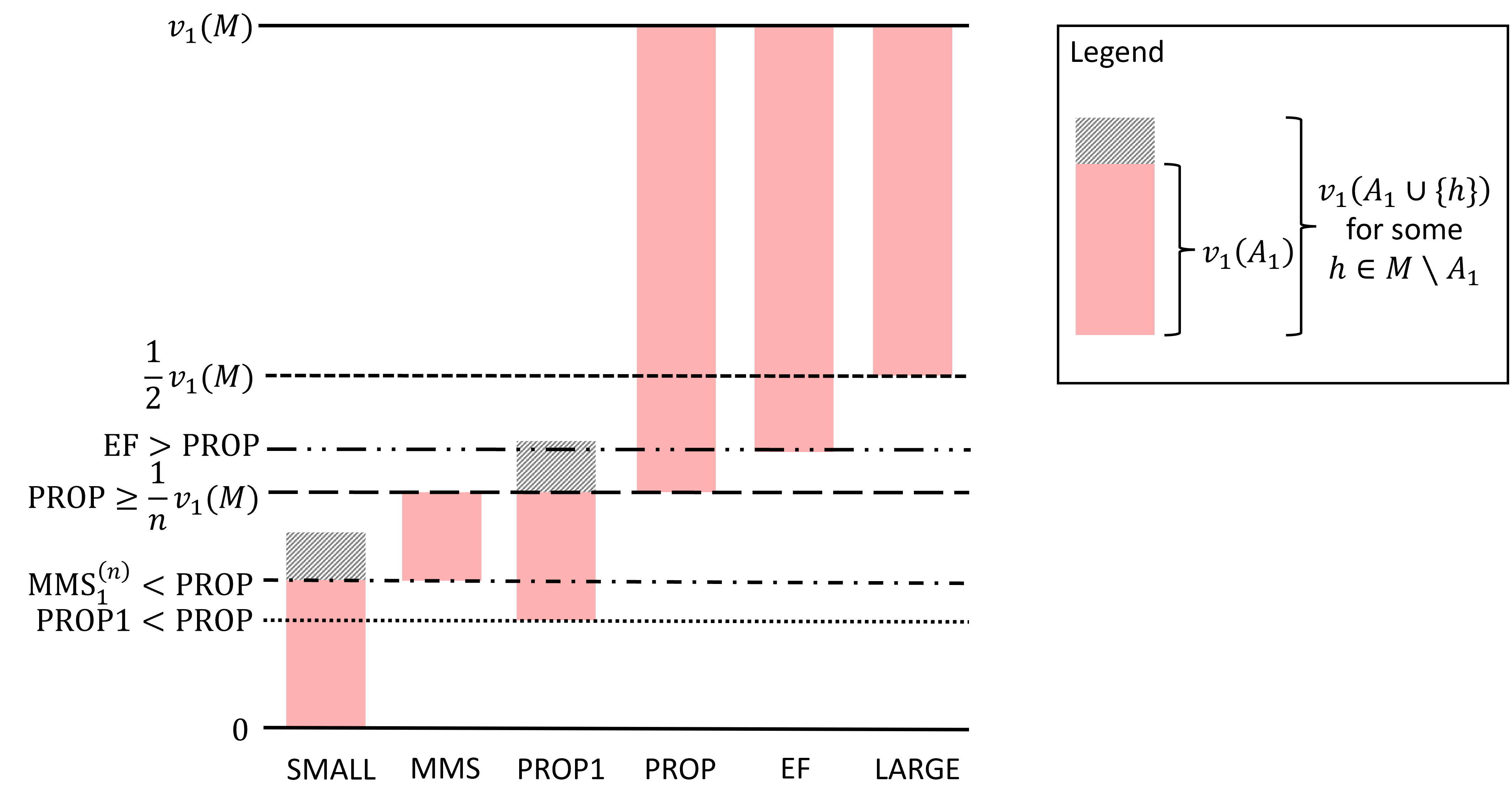}
    \caption{Our selection criteria for allocations of each type which restrict the fractional share of the total value for all resources that the participant can receive. The selection criteria help distinguish between  different types of allocations. The participant's value for their bundle when subjected to each type of allocation is represented by the region shaded in solid red. For example, under \Proptt{}, the participant receives at least $\frac{1}{n}$ of their value for all goods. When subjected to allocations satisfying an approximate fairness notion like \MMStt{}, the value the participant receives is at least the \MMS{} threshold. To avoid confounding factors with a stronger notion like \Prop{}, the value received by the participant is bounded above by the proportionality threshold. For \Proptt{1}, the participant's bundle is valued at under the target proportionality threshold, but adding at most one good to their bundle will take the total value above this target threshold, indicated by the region shaded in gray.}
    \label{fig:thresholds}
\end{figure}

\begin{figure}[t] 
    \centering
    \includegraphics[width=0.55\linewidth]{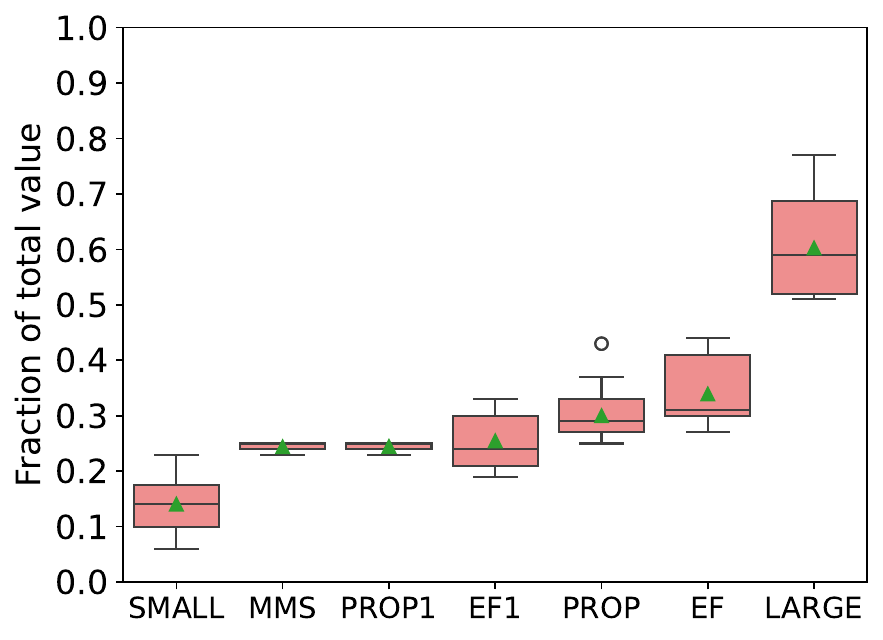}
    \caption{Fraction of the total possible value received by participants in each type of allocation we consider. For each type of allocation, the upper and lower whiskers indicate the maximum and minimum share any participant receives respectively, the upper and lower edges of each box indicate the $75$-th and $25$-th percentile respectively, the horizontal line across the box indicates the median, and the triangular marker shows the mean of the fractional shares received across all participants.    
    }
    \label{fig:actual_fraction}
\end{figure}

In order to compare the effect of the fractional share of resources on perceived fairness, we consider allocations
satisfying one of five 
threshold-based fairness notions. For simplicity, without loss of generality, we will refer to the participant as agent $1$. In our dataset, an allocation $A$ satisfies:
\begin{itemize}
    \item \texttt{\Proptt{}}: if $v_i(A_i) \geq \frac{1}{n} v_i(M)$, for all agents $i \in N$;
    \item \texttt{\Proptt{1}}: if $A$ satisfies \Prop{1} and agent $1$'s bundle does not satisfy \Prop{} (i.e., $v_1(A_1) < \frac{1}{n} v_1(M)$); 
    \item \MMStt{}: if $A$ satisfies \MMS{} and agent $1$'s bundle does not satisfy \Prop{};
    \item \texttt{Small}: $v_1(A_1) <  \MMS{}_1^{(n)}(M)$ and $\exists g \in M \backslash A_1$ such that $v_1(A_1 \cup \{g\}) \geq \MMS{}_1^{(n)}(M)$;
    \item \texttt{Large}: $v_1(A_1) \geq \frac{1}{2} v_1(M)$
\end{itemize}
where we refer to the participant as agent $1$. We compare these five allocation types under treatment T1. We add the restrictions on \Prop{}, \Prop{1}, and \MMS{}, defined in Section \ref{sec:model_defs}, to distinguish these notions from the participant's perspective; otherwise, an allocation $A$ satisfying \Prop{} would entail \Prop{1} and \MMS{}, which may confound our experimental results. \Cref{fig:thresholds} illustrates the fractional share of the total value that the participant may possibly receive when subjected to each type of allocation. \Cref{fig:actual_fraction} describes the share participants actually received under different types of allocations in our experiments. Participant's \texttt{small} bundles are valued at a little under their \MMS{} threshold, while \texttt{large} bundles provide the participant with a majority of the resources.

For all other treatments, we compare \Proptt{}, \Proptt{1}, and \MMStt{}, using these restricted definitions, to the envy-based notions of \EFtt{} and \EFtt{1}. 
To distinguish between \EFtt{} and \EFtt{1}, we assert that \EFtt{1} satisfies \EF{1} and  that agent 1's bundle does not satisfy \EF{} (i.e., $\exists j \in N \backslash \{1\}~:~v_1(A_1) <  v_1(A_j)$). 

\subsection{Survey Outline}
\label{sec:survey_outline}

Participants first gave their consent to our IRB-approved study after being informed of the study description, benefits, risks, rights, and contact information. They were then assigned a treatment and completed a tutorial to introduce them to our framework. Participants subsequently answered ten scenario questions, two of each allocation type.
For each scenario, the allocation was chosen at random from the appropriate data set without replacement, and then the ten scenarios were randomly shuffled.

\vspace{0.5em}
\noindent{\bf Tutorials.}
All participants were required to correctly answer some tutorial questions prior to the scenarios. 
Under the \texttt{\private{}} information treatments, participants were instructed that their bundle's value is ``the sum of values of the items in the treasure.'' Participants received a sample scenario (similar to Figures \ref{fig:sample_scenario_T1} and \ref{fig:sample_scenario_T2_T4}(a)) with their valuation redacted and were asked to compute their bundle's value; see Figures \ref{fig:tutorial_A} and \ref{fig:tutorials_BCDE}(a) in Appendix \ref{apx:tutorials}.

Under \texttt{\public{}} information treatments, participants were taught that bundle values may differ across agents. Specifically, participants were told that their bundle's value ``may differ from how any other pirate values the treasure.'' In each scenario (similar to Figure \ref{fig:sample_scenario_T2_T4}(b)) participants were asked to identify how much (i) they value their own bundle, (ii) Pirate 1 values their own bundle, (iii) Pirate 1 values the participant's bundle; see also Figure \ref{fig:tutorials_BCDE}(b) in Appendix \ref{apx:tutorials}. This comprehension task primed participants to consider other agents' perspectives in their decision-making.

\vspace{0.5em}
\noindent{\bf Self-reported difficulty.}
The ten scenarios were succeeded by a question asking participants to rate the difficulty of the scenarios on a $5$-point Likert scale from Very Easy (1) to Very Hard (5).

\vspace{0.5em}
\noindent{\bf Attentiveness check questions.}
Prior to the tutorial, participants
answered a simple arithmetic problem to ensure they were not bots, which is a known problem for Mechanical Turk \citep{kennedy20:shape}. Furthermore, on the final page, participants reported their favorite
good and final comments or questions. 

\vspace{0.5em}
\noindent{\bf Participant qualifications.}
In order to obtain high-quality responses, participation in our study was restricted to Mechanical Turk workers who (a) had at least an 80\% approval rate on previous tasks, (b) had completed at least 100 tasks, (c) were located in either the United States or Canada,\footnote{We restricted location to ensure language proficiency and prevent any potential issues due to linguistic barriers.} (d) had a Master’s qualification\footnote{Workers with `Master’s qualification' have ``consistently demonstrated a high degree of success in performing a wide range of HITs across a large number of Requesters.'' See \url{https://www.mturk.com/worker/help}.} on the Mechanical Turk platform, and (e) had not attempted or taken the survey before.

\section{Experimental Results}
\label{sec:experiments}

Each participant in our study was first assigned one of the five treatments described in \Cref{tab:treatments}, each of which provides the participant with different amounts of bundle and value information, and a particular fairness measure. They were then presented with ten scenarios. Each scenario consists of an instance of the fair division problem and an allocation that satisfied one of the theoretical fairness criteria, and a question about the participant's perception of their bundle's fairness, as we illustrate in Figures~\ref{fig:sample_scenario_T1} and \ref{fig:sample_scenario_T2_T4}. In this section, we discuss our findings about the three primary research hypotheses discussed in the introduction. We dedicate a separate subsection to each hypothesis and focus our attention at only those parts of our experimental setup that are relevant to the hypothesis being considered. We discuss our analysis of other factors influencing perceived fairness in \Cref{sec:factors_influencing}.


\subsection{(H1) Perceived fairness vs. amount of resources} 
\label{sec:h1_perceived}


Our first study aims to determine the relationship between the amount of resources received by participants and perceived fairness. In each scenario, we offer participants allocations that satisfy one of five threshold-based notions -- \smallqtt{}, \MMStt{}, \Proptt{1}, \Proptt{}, or \largeqtt{} -- and measure perceived fairness explicitly by asking participants whether they find their allocation acceptable. This setup follows treatment T1 by providing participants with information only about their own bundle and their own subjective values for the goods (see Figure \ref{fig:sample_scenario_T1}). 

\bigbreak
\emph{Null hypothesis:} Perceived fairness is independent of allocation fairness property.

\emph{Alternative hypothesis:} Perceived fairness is dependent on allocation fairness property.
\bigbreak

\begin{figure}[t] 
    \centering
     \includegraphics[width=0.65\linewidth]{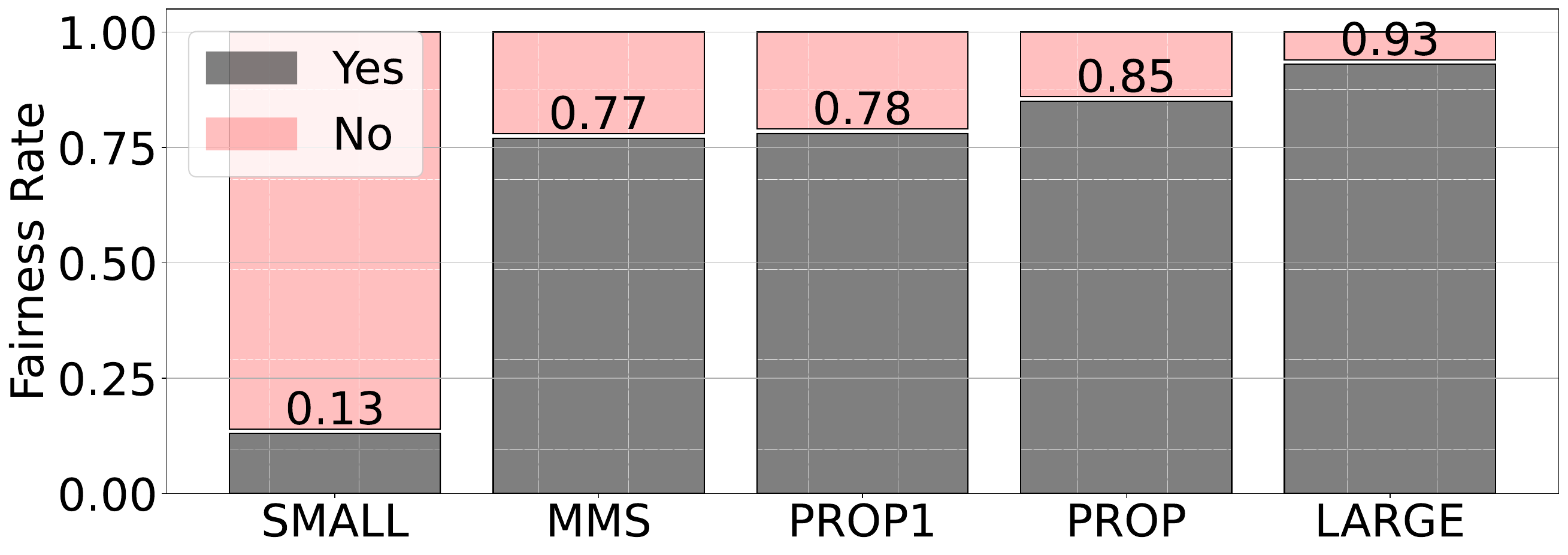}
    \caption{Perceived fairness rate for treatment T1 by fairness type.}
    \label{fig:SR_exp_A}
\end{figure}

The perceived fairness rate -- i.e., fraction of scenarios where participants found their bundle fair -- is demonstrated in Figures \ref{fig:SR_exp_A} and \ref{fig:fairness_heatmap} with column ``T1.'' 
We find significant evidence to reject the null hypothesis that perceived fairness is independent of fairness property between \smallqtt{} and each other property evaluated. The $\chi^2$ test results are presented in Table \ref{tab:swap_rates_by_treatment} under the column ``T1.'' However, there is not enough evidence to suggest that perceived fairness differs between any other pair of fairness properties. 

These tests suggest that there is indeed a relationship between perceived fairness and the amount of resources received. Furthermore, as depicted by Figure \ref{fig:SR_exp_A}, a threshold-based model using the MMS threshold seems to provide significant explanatory power for which type of allocations people deem fair. We explore this possibility further in Section \ref{sec:h4_feature_explain}, below, and discuss other factors that may contribute to participants' decision-making.

\begin{table*}[t]
    \footnotesize{}
    \centering
    \caption{ 
    $\Chi^2$ test statistic and $p$-value for testing the independence of perceived fairness rates under different pairs of treatments, and adjusting for different variables.
    The $\Chi^2$ test is used except when the $p$-value is annotated with a ``$\dagger$'', in which case, it is the result of Fisher's exact test. 
    Higher perceived fairness is indicated with left- or right-arrow.
    }
    \begin{tabular}{|c||c|c|c|c|c|c|c|c|c|}\hline
        \multirow{2}{*}{\bf Pairs of Fairness Notions} & \multicolumn{5}{c|}{\bf Treatment} \\ \cline{2-6}
         & T1 & T2 & T3 & T4 & T5 \\ \hline \hline
        \smallqtt{} vs. \MMStt{} & 
            \makecell{$\chi^2 = 48.6 ~(\rightarrow)$\\\footnotesize{$p < 0.001$}} &
            \cellcolor{lightgray} & 
            \cellcolor{lightgray} & 
            \cellcolor{lightgray} & 
            \cellcolor{lightgray} \\
        \hline
        \smallqtt{} vs. \Proptt{1} & 
            \makecell{$\chi^2 = 51.1 ~(\rightarrow)$\\\footnotesize{$p < 0.001$}} &
            \cellcolor{lightgray} & 
            \cellcolor{lightgray} & 
            \cellcolor{lightgray} & 
            \cellcolor{lightgray} \\
        \hline
        \smallqtt{} vs. \Proptt{} & 
            \makecell{$\chi^2 = 61.7 ~(\rightarrow)$\\\footnotesize{$p < 0.001$}} &
            \cellcolor{lightgray} & 
            \cellcolor{lightgray} & 
            \cellcolor{lightgray} & 
            \cellcolor{lightgray} \\
        \hline
        \smallqtt{} vs. \largeqtt{} & 
            \makecell{$\chi^2 = 77.1 ~(\rightarrow)$\\\footnotesize{$p < 0.001$}} &
            \cellcolor{lightgray} & 
            \cellcolor{lightgray} & 
            \cellcolor{lightgray} & 
            \cellcolor{lightgray} \\
        \hline
        \MMStt{} vs. \Proptt{1} & 
            \makecell{$p: ns$} &
            \makecell{$p: ns$} &
            \makecell{$\dagger p: ns$} & 
            \makecell{$p: ns$} & 
            \makecell{$p: ns$} \\
        \hline
        \MMStt{} vs. \Proptt{} & 
            \makecell{$p: ns$} &
            \makecell{$\chi^2 = 11.6 ~(\rightarrow)$\\\footnotesize{$p < 0.01$}} & 
            \makecell{$\chi^2 = 23.0 ~(\rightarrow)$\\\footnotesize{$p < 0.001$}} & 
            \makecell{$p: ns$} & 
            \makecell{$p: ns$} \\
        \hline
        \MMStt{} vs. \largeqtt{} & 
            \makecell{$p: ns$} &
            \cellcolor{lightgray} & 
            \cellcolor{lightgray} & 
            \cellcolor{lightgray} & 
            \cellcolor{lightgray} \\
        \hline
        \Proptt{1} vs. \Proptt{} & 
            \makecell{$p: ns$} &
            \makecell{$p: ns$} &
            \makecell{$\chi^2 = 18.4 ~(\rightarrow)$\\\footnotesize{$p < 0.001$}} & 
            \makecell{$p: ns$} & 
            \makecell{$p: ns$} \\
        \hline
        \Proptt{1} vs. \largeqtt{} & 
            \makecell{$p: ns$} &
            \cellcolor{lightgray} & 
            \cellcolor{lightgray} & 
            \cellcolor{lightgray} & 
            \cellcolor{lightgray} \\
        \hline
        \Proptt{} vs. \largeqtt{} & 
            \makecell{$p: ns$} &
            \cellcolor{lightgray} & 
            \cellcolor{lightgray} & 
            \cellcolor{lightgray} & 
            \cellcolor{lightgray} \\
        \hline
        \EFtt{} vs. \MMStt{} & 
            \cellcolor{lightgray} & 
            \makecell{$p: ns$} &
            \makecell{$\chi^2 = 86.7 ~(\leftarrow)$\\\footnotesize{$p < 0.001$}} &
            \makecell{$p: ns$} &
            \makecell{$\chi^2 = 33.6 ~(\leftarrow)$\\\footnotesize{$p < 0.001$}} \\
        \hline
        \EFtt{} vs. \Proptt{1} & 
            \cellcolor{lightgray} & 
            \makecell{$p: ns$} & 
            \makecell{$\chi^2 = 80.0 ~(\leftarrow)$\\\footnotesize{$p < 0.001$}} &
            \makecell{$\chi^2 = 8.15 ~(\leftarrow)$\\\footnotesize{$p < 0.05$}} &
            \makecell{$\chi^2 = 31.3 ~(\leftarrow)$\\\footnotesize{$p < 0.001$}} \\
        \hline
        \EFtt{} vs. \Proptt{} & 
            \cellcolor{lightgray} & 
            \makecell{$\dagger p: ns$} &
            \makecell{$\chi^2 = 30.2 ~(\leftarrow)$\\\footnotesize{$p < 0.001$}} &
            \makecell{$p: ns$} &
            \makecell{$\chi^2 = 14.7 ~(\leftarrow)$\\\footnotesize{$p < 0.01$}} \\
        \hline
        \EFtt{} vs. \EFtt{1} & 
            \cellcolor{lightgray} & 
            \makecell{$\chi^2 = 14.4 ~(\leftarrow)$\\\footnotesize{$p < 0.01$}} &
            \makecell{$\chi^2 = 70.8 ~(\leftarrow)$\\\footnotesize{$p < 0.001$}} &
            \makecell{$p: ns$} &
            \makecell{$\chi^2 = 29.0 ~(\leftarrow)$\\\footnotesize{$p < 0.001$}} \\
        \hline
        \EFtt{1} vs. \MMStt{} & 
            \cellcolor{lightgray} & 
            \makecell{$p: ns$} &
            \makecell{$p: ns$} &
            \makecell{$p: ns$} &
            \makecell{$p: ns$} \\
        \hline
        \EFtt{1} vs. \Proptt{1} & 
            \cellcolor{lightgray} & 
            \makecell{$\chi^2 = 9.40 ~(\rightarrow)$\\\footnotesize{$p < 0.05$}} &
            \makecell{$p: ns$} &
            \makecell{$p: ns$} &
            \makecell{$p: ns$} \\
        \hline
        \EFtt{1} vs. \Proptt{} & 
            \cellcolor{lightgray} & 
            \makecell{$\chi^2 = 23.8 ~(\rightarrow)$\\\footnotesize{$p < 0.001$}} &
            \makecell{$\chi^2 = 12.9 ~(\rightarrow)$\\\footnotesize{$p < 0.01$}} &
            \makecell{$p: ns$} &
            \makecell{$p: ns$} \\
        \hline 
    \end{tabular}
    \label{tab:swap_rates_by_treatment}
\end{table*}

\subsection{(H2) Relative perceptions of theoretical fairness notions} 
\label{sec:h2_threshold_vs_envy}


Our second study aims to compare perceptions of fairness across threshold-based and envy-based notions. As with Section \ref{sec:h1_perceived}, we determine whether perceived fairness significantly depends on the fairness property of their provided allocation and expand our analysis to include other treatments which subject the participants to different amounts of bundle and value information and evaluate different measures of fairness. 
Specifically, we test the following hypothesis. 

\bigbreak
\emph{Null hypothesis:} Perceived fairness is independent of allocation fairness property.

\emph{Alternate hypothesis:} Perceived fairness depends on allocation fairness property.
\bigbreak


The perceived fairness rate for each treatment and allocation fairness property is demonstrated in Figure \ref{fig:fairness_heatmap} and the results of $\chi^2$ tests are presented in Table \ref{tab:swap_rates_by_treatment}. We find mixed evidence for rejecting the null hypothesis that perceived fairness is independent of fairness property. In particular, \EFtt{} and \Proptt{} consistently appear as the fairest, across all treatments, followed by their respective relaxations: \Proptt{1}, \MMStt{}, and \EFtt{1}. Participants consistently find allocations satisfying \EFtt{} more fair than those satisfying \Proptt{}, among implicit treatments (T3 and T5), but there is no significant difference among explicit treatments (T2 and T4). 
Hence, while participants in our study prefer the stricter notions of fairness, we do not find enough evidence to suggest threshold-based notions are more fair than envy-based ones.

\subsection{(H3) Sensitivity Analysis} 
\label{sec:h3_sensitivity_analysis}

\begin{figure}[t]
    \centering
    \includegraphics[width=0.75\linewidth]{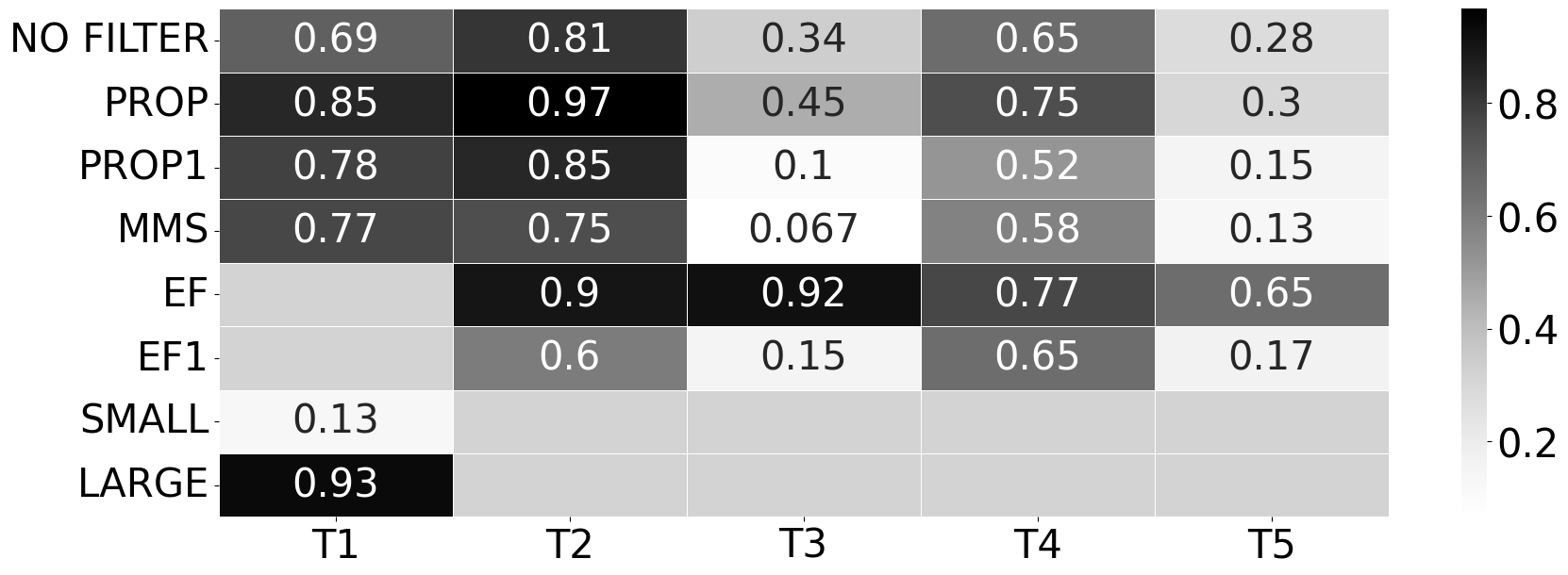}
    \caption[]{Heat map of perceived fairness per treatment, across different types of allocations. The value in each cell is the fraction of participants who perceived their bundle to be fair. A larger value corresponds to greater perceived fairness.\footnotemark 
    
    }
    \label{fig:fairness_heatmap}
\end{figure}

\footnotetext{For treatments T1, T2, and T4, this number denotes the fraction of responses where the participant clicked ``Yes'' when asked ``Do you find your share acceptable?'' For T3 and T5, this number denotes the fraction of responses where the participant chose \textit{not} to swap, when asked ``Would you like to keep the treasure assigned to you or swap with one of the other pirates?''}

Recall that both threshold- and envy-based theories of fairness are \emph{intrapersonal}, meaning that a person's assessment of distributive fairness does not depend on interpersonal comparisons of utility \citep{sen1999,robbins1938interpersonal}. If true, this would entail that perceived fairness should not depend on what information about other agents' bundles or values is conveyed to participants. 
Our third study is therefore a sensitivity analysis of our previous findings from Section \ref{sec:h2_threshold_vs_envy} against different treatments of contextual framing \citep{tversky1985framing}. In particular, we study the effects of bundle information, value information, and fairness measure on perceptions of fairness. 
We formalize our research question and statistical hypothesis as follows.

\bigbreak
\emph{Research Question:} For any pair of treatments $X$ and $Y$ that is different on a single axis (bundle information, value information, or fairness measure), does perceived fairness differ between $X$ and $Y$ overall and when controlling independently for allocation fairness property?

\emph{Null hypothesis:} Perceived fairness is independent of treatment.

\emph{Alternate hypothesis:} Perceived fairness depends on treatment.
\bigbreak

Figure \ref{fig:fairness_heatmap} presents a heat map of perceived fairness rates across treatments and by allocation fairness properties.
From the top row labeled ``All Scenarios,'' we observe that participants perceive their bundle more fairly as bundle information is increased (i.e., T2 $\succ$ T1), value information is more private (i.e., T2 $\succ$ T4 and T3 $\succ$ T5), and participants are asked about fairness explicitly rather than implicitly (i.e., T2 $\succ$ T3 and T4 $\succ$ T5). 
Our experiments provide statistically significant evidence for rejecting the null hypothesis that perceived fairness is independent of treatment. We draw this conclusion using the $\chi^2$ test with $p < 0.01$ for nearly all treatment pairs, except for the pair of (T3, T5).
This is demonstrated by Table \ref{tab:swap_rates} in Appendix \ref{app:sensitivity}. 

Our results demonstrate that perceived fairness indeed depends on experimental framing, thus confirming recent evidence that perceptions of algorithmic decisions are context-dependent \citep{lee18:understanding}. This provides evidence to suggest that individuals' understanding of distributive fairness is not entirely interpersonal, as theorized by threshold- and envy-based doctrine.







%

\section{Factors Influencing Perceived Fairness}\label{sec:factors_influencing}

Based on our finding that perceived fairness is influenced by experimental framing (\Cref{sec:h3_sensitivity_analysis}), we further examine the criteria participants use to decide whether an allocation is fair. This is done by measuring i) the relative influence of various characteristics of the resource allocation instances presented, ii) the presence of various motivations behind participants' decisions to swap their bundle, and iii) the differences in cognitive effort exerted by participants across different treatments.


\subsection{Predicting and Explaining Perceived Fairness}
\label{sec:h4_feature_explain}



We conduct an analysis of the influence different properties of an allocation have on the participants' decisions about whether the allocation is fair, in each of the five treatments. To this end, we fit machine learning (ML) models such as Logistic Regression (LR) and Decision Tree (DT)-based models, including Random Forest (RF) and XGBoost (XGB) \citep{chen2016xgboost}, on a set of features that is determined by the amount of information available to participants in the corresponding treatment. For example:
\begin{itemize}
    \item \textit{IsMMS} is a binary indicator for whether the participant's value is greater than their MMS threshold, i.e. $\mathds{1}\{v_1(A_1) \geq MMS_1^{(n)}(M)\}$.
    \item \textit{FracValue} is the value the participant has for their bundle as divided by their total value for all goods, i.e. $v_1(A_1)/v_1(M)$.  
    \item \textit{BestSelfImprovement} is the maximum gain the participant can experience by swapping their bundle (negative if not swapping is optimal), i.e. $\max_{k \in N \setminus \{1\}} v_1(A_k) - v_1(A_1)$.
\end{itemize}
The full list of features considered for each treatment is provided in  \Cref{tab:factors_list} in \Cref{app:feature_imp}.



\begin{table}[t]\centering
\scriptsize
\caption{Three most important features, according to the Logistic Regression model, in terms of feature coefficients (absolute value is indicated in brackets).
}
\resizebox{0.8\columnwidth}{!}{
\begin{tabular}{llllll}\toprule
\textbf{T1} &\textbf{T2} &\textbf{T3} &\textbf{T4} &\textbf{T5} \\\midrule
IsMMS (2.55) &IsMMS (1.6) &BestSelfImprovement (3.04) &IsProp (0.97) &BestSelfImprovement (1.11) \\
GetsHighest (0.54) &IsProp (1.0) &MeanSelf (1.75) &FracValue (0.80) &FracValue (0.96) \\
NumItems (0.46) &StdSelf (0.72) &FracValue (1.41) &IsEF (0.77) &IsEF (0.95) \\
\bottomrule
\end{tabular}}
\label{tab:lr_features}
\end{table}

\begin{table}[t]
\centering
\scriptsize
\caption{Three most important features, according to the  Decision Tree model, in terms of the Gini index (indicated in brackets).}
\resizebox{\columnwidth}{!}{
\begin{tabular}{llllll}\toprule
\textbf{T1} &\textbf{T2} &\textbf{T3} &\textbf{T4} &\textbf{T5} \\\midrule
FracValue (0.96) &FracValue (0.61) &BestSelfImprovement (0.87) &FracValue (0.28) &BestSelfImprovement (0.63) \\
GetsHighest (0.02) &MeanSelf (0.15) &MeanSelf (0.04) &BestNetImprovement (0.13) &StdSelf (0.07) \\
NumItems (0.02) &MaxSelf (0.09) &MaxSelf (0.03) &MeanSelf (0.11) &MinAll (0.06) \\
\bottomrule
\end{tabular}}
\label{tab:dt_features}
\end{table}

\Cref{tab:lr_features} and \Cref{tab:dt_features} indicate the most important features (and their relative influence) as per LR and DT, in each treatment.\footnote{See \cref{app:feature_imp} for details about RF and XGB, which give results similar to DT.} For LR, the most important features are obtained by comparing the absolute values of the weights learned, while the Gini index is used to measure relative importance in the case of DT-based models.\footnote{A higher feature weight learned by LR indicates a greater influence of that feature on the target variable. Similarly, Gini index measures how effective a feature is in splitting the data in terms of the target variable (larger values imply greater influence).} We find that features such as \textit{IsMMS, IsProp,} and \textit{FracValue}, which consider only the participant's assessment of their \textit{own} bundle, are the most influential with the \textit{threshold-based} (implicit) fairness measure. On the other hand, all models agree that the most important feature in treatments with the \textit{envy-based} (explicit) fairness measure, is \textit{BestSelfImprovement}, which considers the participant's assessment of their bundle \textit{relative} to others' bundles. Hence, the \textit{share} received by participants is more predictive of perceived fairness in treatments with the threshold-based fairness measure, while a \textit{comparison} between bundles is more informative in treatments where the fairness measure is envy-based. 
In \Cref{app:feature_imp} we show that the models fit on such features yield significant improvements, in terms of predicting participants' perception of fairness, over a baseline that always predicts the majority response for each treatment. 

\subsection{Selfishness Motivates Swaps}
\label{sec:movtivating_swaps}

In Section \ref{sec:h3_sensitivity_analysis}, we demonstrate how perceptions of fairness can be influenced by question framing---overall, fewer participants find the allocation fair if asked whether they wish to \textit{swap}, as opposed to being asked if the allocation is \textit{acceptable}. In this section, we analyze the relative influence of different possible motivations behind participants’ swaps, such as \textit{selfishness}, \textit{altruism}, and a concern for \textit{economic efficiency}. We also study the effect of providing information about other agents’ valuations (T5) as compared to the case when subjects are not aware that other agents can have valuations different from their own (T3).

\vspace{0.5cm}
\noindent{\bf Selfishness vs. Altruism.}
We consider a swap to be selfish if the value received by the participant (agent $1$) is increased over the default and the agent ($j$) they swap with ends up with a lower value. Analogously, we consider a swap as altruistic if the value received by another agent is increased at the participant’s own cost. Note that in T3, participants are likely to assume that other agents have the same valuation as them, since no information (such as that in T5) is provided about other agents' values. To account for this, we use the definitions provided in \Cref{tab:swap_definitions}, which consider only the participant's valuation in T3, but both the participant's and agent $j$'s valuation in T5. \Cref{tab:selfish_altruistic} summarizes participants’ behavior with respect to selfish and altruistic swaps.

\begin{table}[t]
    \centering
    \caption{Conditions for selfish and altruistic swaps in T3 and T5. The participant is represented by agent $1$ and agent $j$ is the agent they swap with. $A_1$ and $A_j$ are their respective original bundles.}
    \begin{tabular}{|c|c|c|}\hline
     \textbf{Type of Swap} & \textbf{T3} & \textbf{T5} \\\hline
      Selfish   & $v_1(A_j) > v_1(A_1)$ & $v_1(A_j) > v_1(A_1) \land v_j(A_j) > v_j(A_1)$ \\
      Altruistic & $v_1(A_1) > v_1(A_j)$ & $v_1(A_1) > v_1(A_j) \land v_j(A_1) > v_j(A_j)$\\
      \hline
    \end{tabular}
    \label{tab:swap_definitions}
\end{table}


We observe that in both treatments, participants opt for selfish swaps in a large percentage of instances where such swaps are possible. A much smaller percentage of altruistic swaps is observed, indicating that selfishness is a stronger motivation for swapping as compared to altruism. Interestingly, the fraction of selfish swaps is significantly lower in T5 than in T3 (at $p < 0.005$), indicating that more value information leads to a decrease in selfish behavior. On the other hand, the increase in the fraction of altruistic swaps in T5 (compared to T3) is statistically significant (at $p < 0.001$). Yet, due to the small sample size (number of altruistic swaps possible) in T5, we do not make a stronger claim about the relationship between altruism and value information. Given the observations of \citet{locey2015altruism}, that people behave more altruistically when they have more information about each other, this remains an interesting direction for further investigation.

\begin{table}[t] 
\centering
\scriptsize
\caption{Rates of selfish and altruistic swaps in treatments T3 and T5. The significance level for changes across both treatments is computed using Fisher's exact test. 
}
\begin{tabular}{cccccccccc}\toprule
&\multicolumn{3}{c}{\textbf{T3}} & &\multicolumn{3}{c}{\textbf{T5}} &  \\\cmidrule{2-4}\cmidrule{6-8}
\textbf{Type of swap} &\textbf{\makecell{Swaps\\made}} &\textbf{\makecell{Swaps\\possible}} &\textbf{\makecell{Percentage\\utilized}} & &\textbf{\makecell{Swaps\\made}} &\textbf{\makecell{Swaps\\possible}} &\textbf{\makecell{Percentage\\utilized}} &\textbf{\makecell{Significant\\difference}} \\\midrule
\textbf{Selfish} &190 &224 &\textbf{84.82} & &134 &187 &\textbf{71.65} &$\boldsymbol{p < 0.005}$ \\
\textbf{Altruistic} &8 &289 &2.77 & &7 &41 &17.07 &$\boldsymbol{p < 0.001}$ \\
\bottomrule
\end{tabular}
\label{tab:selfish_altruistic}
\end{table}

\vspace{0.5cm}
\noindent{\bf Economic Efficiency.}
Since there appears to be an effect of value information on different types of swaps, we next examine whether the knowledge of other agents’ valuations motivates participants to make swaps that improve the \textit{overall good}, a.k.a. \textit{welfare}. For this, we analyze two types of swaps that improve the efficiency of the allocation:
\begin{itemize}[leftmargin=*]
    \item \emph{Pareto improvement}: there is an increase in the value received by both the participant and agent $j$, i.e. when $v_1(A_j) > v_1(A_1)$ and $v_j(A_1) > v_j(A_j)$;
    \item \emph{utility-maximizing}: the sum of the values received by the participant and agent $j$ increases, i.e. when $v_1(A_j) + v_j(A_1) > v_1(A_1) + v_j(A_j)$.
\end{itemize}
Note that since the participants do not have access to the valuations of other agents in T3, any Pareto improvement or utility-maximizing swaps are more likely by chance than by intention. We observe that there is no significant increase (as per Fisher's exact test \citep{kim2017statistical}) in the fraction of either of these swaps in T5, which indicates that efficiency is not a clear motivation for swaps, when participants are indeed aware of others’ valuations (see \Cref{tab:pareto_utility} in \Cref{app:motivating_swaps} for details).

\subsection{Cognitive Effort} \label{sec:cognitive}




Our final study is one of the cognitive effort exerted by participants. 
It is known that people conserve cognitive effort because it imposes a cost besides their monetary payoff \citep{westbrook2013subjective}. However, it is not clear whether people will exert significant effort to verify fair outcomes, as found in public goods games by \citet{fehr2002altruistic}.
Here, we focus on the effects of bundle information, value information, and fairness measure on cognitive effort along two dimensions:
\begin{itemize}[leftmargin=1em]
    \item \emph{Response time:} the time in seconds it took participants to answer scenario questions, and
    \item \emph{Reported difficulty:} participants' reported difficulty on a $5$-point Likert scale according to a question conducted at the end of the questionnaire.
\end{itemize}


\bigbreak
\emph{Null hypothesis:} Cognitive effort does not depend on treatment.

\emph{Alternate hypothesis:} Cognitive effort depends on treatment.
\bigbreak


\begin{figure}[t]
\centering
\begin{tabular}{cc}
    \includegraphics[width=0.47\linewidth]{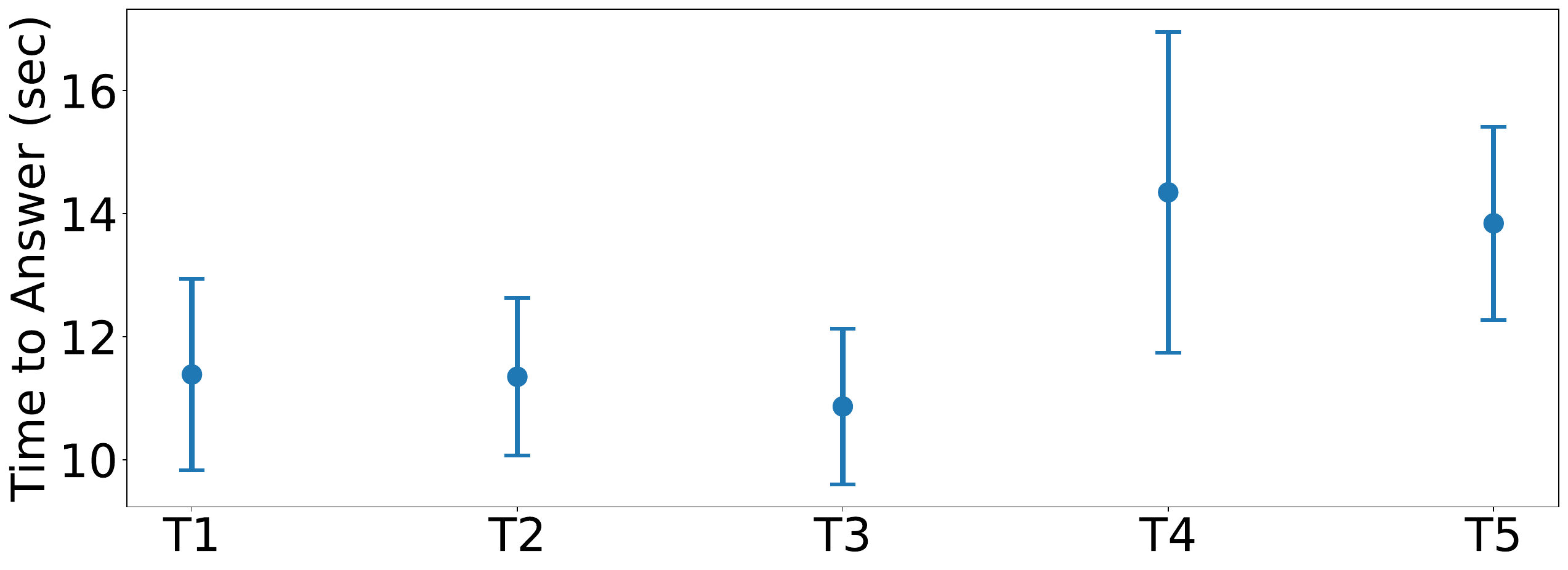}
     & 
    \includegraphics[width=0.47\linewidth]{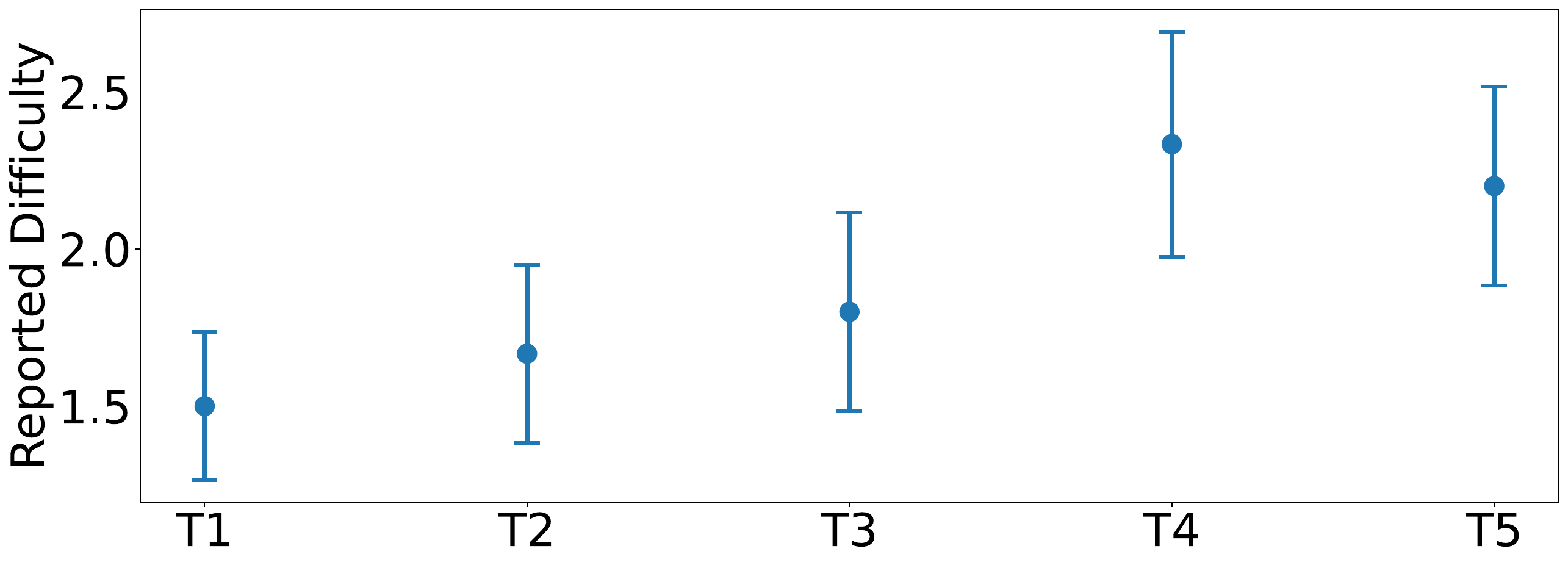}
     \\
    (a) Response time & (b) Reported difficulty
\end{tabular}
\caption{ 
    Means and $95\%$ confidence intervals of cognitive difficulty by treatment, as measured by (a) time to answer (seconds) and (b) reported difficulty ($5$-point Likert scale).
}
\label{fig:cog_diff_figs}
\end{figure} 

Figures \ref{fig:cog_diff_figs}(a) and (b) demonstrate our results for response time and reported difficulty, respectively, while Tables \ref{tab:time_to_answer} and \ref{tab:feedback} in Appendix \ref{apx:cog_effort_charts} demonstrate pairwise $t$-tests in response time and reported difficulty between the five treatments.
We observe some evidence to suggest that cognitive effort depends on value information: it increases as participants are informed of other agents' values. This is evidenced by our finding that almost all pairwise tests between treatments with private and public value information are statistically significant with $p < 0.05$, 
except for the pairs of treatments (T1, T2) in response times (Table \ref{tab:time_to_answer}) and (T3, T5) in reported difficulty (Table \ref{tab:feedback}).  
A few participants articulated their cognitive difficulty in the public value information treatments through their feedback
(see Section \ref{sec:comments}). These include \emph{``I had to think about how much weight to put on each factor''} (T4\#6), \emph{``Some questions need more thinking and attention''} (T4\#19, T4\#29), and \emph{``I felt rushed''} (T4\#4, T5\#27).\footnote{Participant comments are uniquely identified by an index within their treatment -- e.g., T4\#6 is the 6th participant within T4.}
From our findings in Section \ref{sec:h3_sensitivity_analysis}, and as seen in Figure \ref{fig:fairness_heatmap}, this suggests a negative correlation between cognitive effort and perceived fairness. That is, the harder the problem complexity, the more cognitive effort participants exert to evaluate fairness and the more they realize reasons that their bundle may not be fair. This is an interesting finding that we plan to investigate in future work.



We do not find enough evidence to suggest that cognitive effort depends on bundle information.
This is evidenced by our finding that tests between (T1, T2) and (T1, T3) are not significant for either time to answer or reported feedback.
Furthermore, we find little effect in cognitive effort due fairness measure.
This is exemplified in Tables \ref{tab:time_to_answer} and \ref{tab:feedback}, since measures of answer time and reported feedback are not significant between (T2, T3) and (T4, T5). Overall, based on the above observations, we do not find sufficient evidence to reject the null hypothesis (see \Cref{apx:cog_effort_charts} for further details).


\section{Discussion}

In this work, we presented a human subject experiment on Amazon Mechanical Turk testing perceptions of fairness of threshold- and envy-based notions. Participants took the role of one agent in a fair division instance where they were offered an allocation that satisfied one of several theoretical fairness notions. After participants received some information about the allocation and agents' values, they reported whether they perceived their bundle to be fair in a manner dependent on their treatment.
This work contributes to an empirical line of analysis about which theoretical fairness notions are actually representative of human values and decision-making. Our primary contributions are three-fold and provide insight into when people find allocations fair.

We first compared the efficacy of threshold-based notions, \Prop{} and its relaxations \Prop{1} and \MMS{}, by providing participants with bundles whose value exceeded the corresponding thresholds. While providing a greater share of resources does increase perceived fairness, we found that the differences in the rates at which allocations satisfying \Prop{1} is not statistically significant compared to \MMS{} and \Prop{} which guarantee a higher threshold.
Since \Prop{} and \MMS{} allocations may not exist, this suggests that \Prop{1} allocations may be the most useful in practice, as they always exist and can be computed efficiently \citep{conitzer2017fair}.

Our second study compare different threshold-based and envy-based fairness notions and seeks to establish a relationship between them. Our experiments provide empirical evidence that people do in fact perceive the strict fairness notions \Prop{} and \EF{} as fairer than their relaxations \Prop{}, \MMS{}, and \EF{}), as conjectured by theorists. However, there is no evidence to suggest that there is a clear human preference for either threshold-based or envy-based notions.


Our third study shows that human perceptions of fairness are affected by both context and framing, and contributes fresh insights into design implications for practitioners seeking buy-in from participants in resource allocation markets. We found that the method by which perceived fairness is measured significantly affected our results. In particular, the perception of fairness drops when people are offered the opportunity to swap their bundle with another agent. This is supported by our experimental results that the rate at which participants find their bundle to be acceptable is significantly higher than the rate at which they stay with the bundle provided to them when they are allowed to swap.

An important question for future research is 
whether fairness is in fact multi-dimensional and our two measures capture different aspects of fairness, or if the measure itself caused participants to perceive their bundle differently.
In Section \ref{sec:h3_sensitivity_analysis}, we also found that perceived fairness increases with more bundle information, but decreases with information about others' values. These findings confirm those of \citet[Figure 3]{hosseini2022hide} in that people perceive their bundle as fairer if they are more certain about the entire allocation. One implication is that \emph{epistemic envy-freeness}, as introduced by \citet{aziz2018knowledge}, may be perceived as less fair than other relaxations of envy-freeness. Our findings also align with experiments by \citet{locey2015altruism} in that, as people know less personal information about others, they may experience less discomfort by comparing their own bundle to others. This may drive people to behave more conservatively and optimistically about their own resources.





\subsection{Descriptive Comments from Participants}
\label{sec:comments}

Following the questionnaire we asked participants for their comments about the scenarios. 
Many of the participants' comments describe what heuristics they used to answer the scenarios. For example, one participant conveyed that they \emph{``tended to choose the max value for me''} (T5\#30) while another made decisions based on \emph{``how much I valued the items''} and \emph{``my gut of what I thought would be best''} (T5\#22). One participant identified the proportionality rule as their criterion for acceptability: \emph{``as long as I received more than 25\% of the total, I was satisfied (assumed it was fair)''} (T1\#20).

While these heuristics only depend on participants' own bundles, many participants considered others' bundles when making their decisions. For instance, several participants were concerned with equitable treatment. One participant commented that they took decisions \emph{``according that each one have to got the equal share of the treasure''} (T1\#15) while another found the allocations unacceptable if \emph{``there was a large disparity''} (T2\#3). Rather, these participants were \emph{``aiming for more equal allotment''} (T2\#13).
Meanwhile, other participants contrasted what they received with respect to other agents' bundles. Several participants commented that, while their own bundle was acceptable, \emph{``it doesn't necessarily mean that it was fair for the whole crew!''} (T1\#24, T2\#18, T2\#29). One participant therefore \emph{``went by how the others valued my share''} (T4\#1). Another participant took a more complex decision procedure: \emph{``I tried to trade with pirates who offered me a fair price or a discount, but also asked for a fair price for their items''} (T5\#14).

A few participants commented about factors beyond the bundles' monetary value, such as desiring \emph{``a lower value treasure if it has items you like''} (T3\#7) and wondering \emph{``what the future values would be of the items''} (T3\#15). An interesting direction for future work is to study the effects of the interest stakeholders have in the outcome or in receiving highly desirable items on behaviors such as cooperation, altruism and equitability.




\subsection{Limitations and Future Work}
\label{sec:limitations}

There are a number of limitations inherent in our experiment that could be analyzed in further detail with potential follow-up work. As with many economic lab experiments, we were limited by the size, scope, and scale of our study. In particular, scenarios in our study consisted of only four agents, three of which were fictional, and ten items, which did not directly impact participants' welfare outside our questionnaire's context. 

Participants in our study were offered $\$1.00$, independent of their choices, to incentivize truthful responses.
Although there is some evidence suggesting that the size of extrinsic incentives doesn't affect experimental responses \citep{cameron1999raising}, it remains to be seen whether our results would generalize to larger settings. 
Still, our findings do provide statistically significant insight into perceptions of fairness of different distributions of resources. Future work could analyze real-world case studies to determine if people's perception of resource allocations aligns with theory and our experiments.

The sensitivity analysis in Section \ref{sec:h3_sensitivity_analysis} of perceived fairness along the dimensions of bundle information, value information and fairness measure only tests two natural options on the scale of granularity along these dimensions. Future work may expand upon our present results along these or other dimensions.

Finally, we made a standard microeconomic assumption in our analysis that peoples' utility is additive, and only depends on their payoff, as presented in the allocations, rather than a composite of all agents' payoffs or other extrinsic factors. Our analysis of fairness also assumes that agents are a priori equal, as opposed to having intrinsic differences, such as prior wealth or merit, that may affect peoples' understanding of fairness \citep{michelbach2003doing,cetre2019preferences}. Future work may conduct experiments that challenge these assumptions and elicit preferences from human subjects.
It may further be useful to compare perceptions of fairness from people across different cultural backgrounds \citep{henrich2010weirdest}.

\section*{Acknowledgements}

We thank the anonymous reviewers for helpful comments. HH acknowledges support from NSF grants
\#2144413, \#2107173, and \#2052488. LX acknowledges support from NSF grants \#1453542, \#2007994, and
\#2106983, and a Google Research Award.

\clearpage

\bibliographystyle{plainnat}
\bibliography{for_arxiv}

\clearpage




\newpage
\section*{Appendix}

\section{Tutorials}
\label{apx:tutorials}

\begin{figure}[h]
    \centering
    \includegraphics[width=0.45\linewidth]{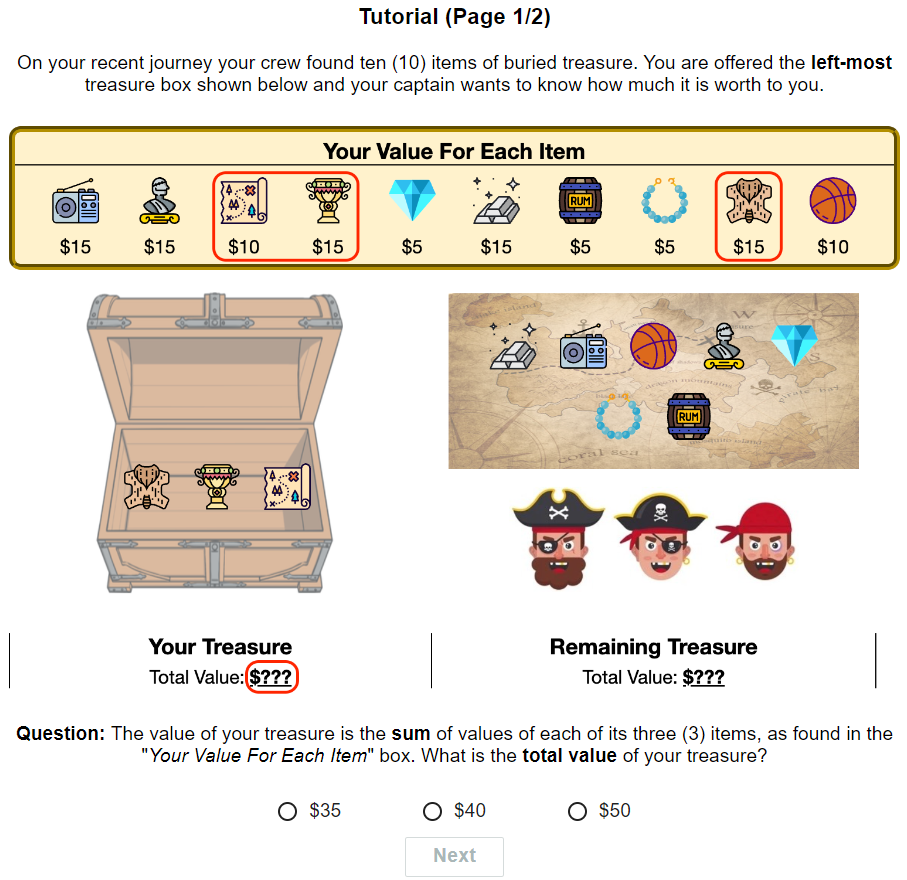}
    \caption{Tutorial for treatment T1.}
    \label{fig:tutorial_A}
\end{figure}

\begin{figure}[h]
\centering
\begin{tabular}{cc}
    \includegraphics[width=0.47\textwidth]{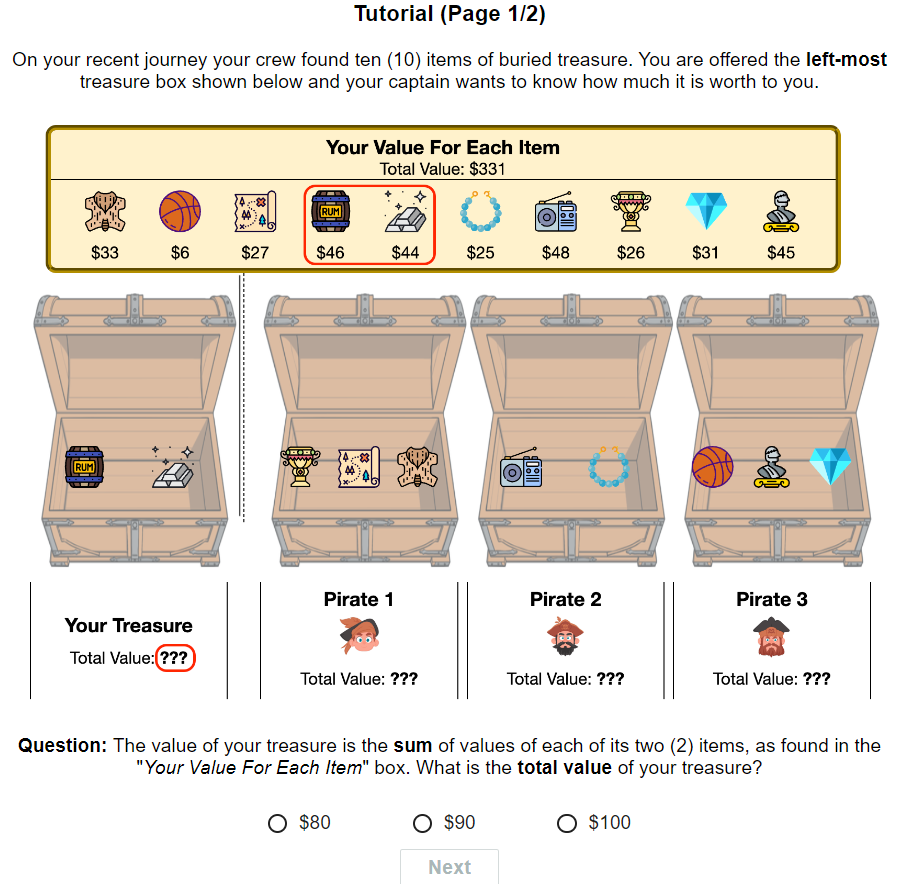}
     & 
    \includegraphics[width=0.47\textwidth]{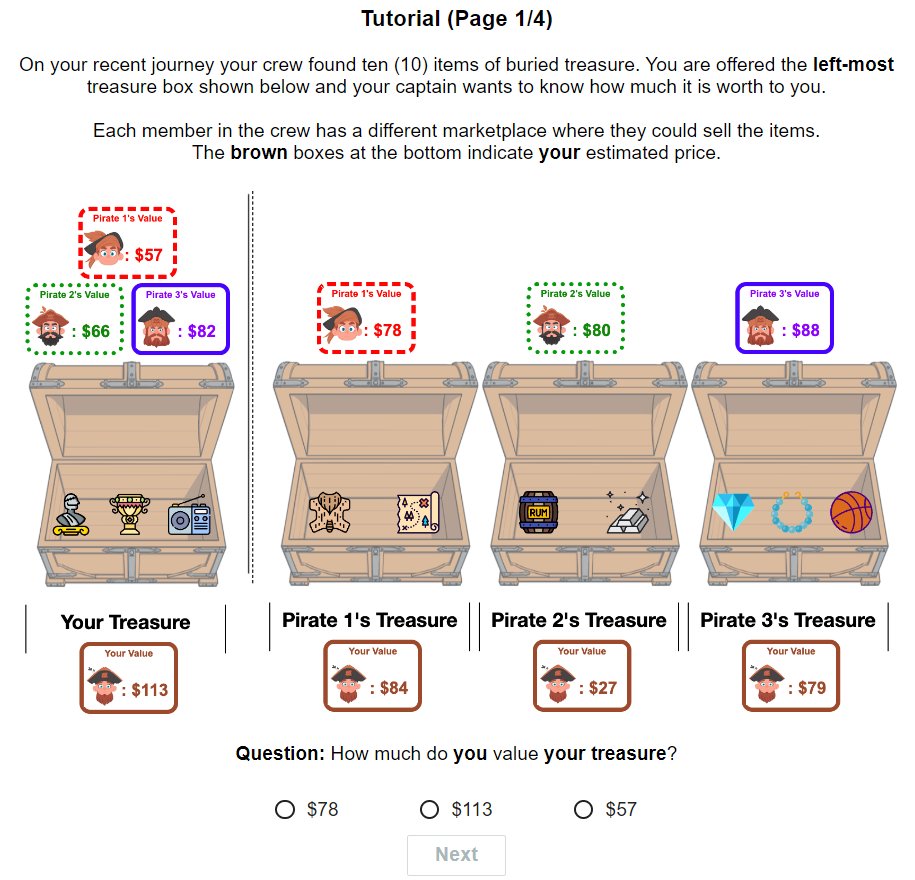}
     \\
    (a) & (b)
\end{tabular}
\caption{ 
    Tutorials for treatments (a) T2 and T3, and (b) T4 and T5.
}
\label{fig:tutorials_BCDE}
\end{figure}

As discussed in Section \ref{sec:survey_outline}, every participant completed some tutorial questions prior to beginning the questionnaire. Participants under different treatments received slightly different tutorial pages, depending on what information they would receive in the subsequent scenarios. For example, one tutorial page for treatment T1 may be found in Figure \ref{fig:tutorial_A}. This taught each participant that their value for their bundle is the sum of the values of items in that bundle. This concept was also taught in the tutorials for treatments T2 and T3, Figure \ref{fig:tutorials_BCDE}(a), except for bundled bundle information and private value information. The tutorial for treatments T4 and T5, in Figure \ref{fig:tutorials_BCDE}(b), instead explicitly taught each participant that their value for each item may differ from the other agents.


\clearpage
\section{Sensitivity to Treatment}\label{app:sensitivity}

\Cref{tab:swap_rates} describes the influence of treatment on perceived fairness. There are significant differences in levels of perceived fairness in all pairs of treatments (except T3 and T5) that vary only in one dimension (i.e. bundle information, value information, or fairness measure). In most cases, this difference also holds when subsets of instances are considered that satisfy specific properties (such as notions of fairness).

\begin{table*}[h]
    \tiny
    \centering
    \caption{ $\Chi^2$ test statistic and $p$-value for testing the independence of perceived fairness rates under different pairs of treatments, and adjusting for different variables.
    Higher perceived fairness is indicated with left- or right-arrow.
    The $p$-value is represented as: a cell labeled \ns{} (not significant) implies that $p\geq 0.05$.
    }
    \begin{tabular}{|c|c||c|c|c|c|c|c|c|c|c|}\hline
        \multirow{2}{*}{\bf Variable} & \multirow{2}{*}{\bf Value} & \multicolumn{5}{c|}{\bf Pairs of Treatments} \\ \cline{3-7}
         & & T1~ vs. T2 & T2~ vs. T4 & T3~ vs. T5 & T2~ vs. T3 & T4~ vs. T5 \\ \hline \hline
        \multicolumn{2}{|c||}{All scenarios} & 
            \makecell{$\chi^2 = 11.6 ~(\rightarrow)$\\\footnotesize{$p < 0.01$}} &
            \makecell{$\chi^2 = 19.6 ~(\leftarrow)$\\\footnotesize{$p < 0.001$}} & 
            \makecell{$p: ns$} & 
            \makecell{$\chi^2 = 139.5 ~(\leftarrow)$\\\footnotesize{$p < 0.001$}} & 
            \makecell{$\chi^2 = 84.0 ~(\leftarrow)$\\\footnotesize{$p < 0.001$}} \\
        \hline
        \multirow{5}{*}{\makecell{Allocation\\ Fairness \\ Property}} 
        & \Proptt{} &
            \makecell{$p: ns$} & 
            \makecell{$\chi^2 = 11.6 ~(\leftarrow)$\\\footnotesize{$p < 0.01$}} & 
            \makecell{$p: ns$} & 
            \makecell{$\chi^2 = 38.7 ~(\leftarrow)$\\\footnotesize{$p < 0.001$}} & 
            \makecell{$\chi^2 = 24.3 ~(\leftarrow)$\\\footnotesize{$p < 0.001$}} \\
        \cline{2-7}
        & \Proptt{1} &
            \makecell{$p: ns$} & 
            \makecell{$\chi^2 = 15.4 ~(\leftarrow)$\\\footnotesize{$p < 0.01$}} & 
            \makecell{$p: ns$} & 
            \makecell{$\chi^2 = 67.7 ~(\leftarrow)$\\\footnotesize{$p < 0.001$}} & 
            \makecell{$\chi^2 = 18.2 ~(\leftarrow)$\\\footnotesize{$p < 0.001$}} \\
        \cline{2-7}
        & \MMStt{} &
            \makecell{$p: ns$} & 
            \makecell{$p: ns$} & 
            \makecell{$p: ns$} & 
            \makecell{$\chi^2 = 58.0 ~(\leftarrow)$\\\footnotesize{$p < 0.001$}} & 
            \makecell{$\chi^2 = 26.4 ~(\leftarrow)$\\\footnotesize{$p < 0.001$}} \\
        \cline{2-7}
        & \EFtt{} &
            \makecell{\texttt{N/A}} & 
            \makecell{$p: ns$} & 
            \makecell{$\chi^2 = 12.6 ~(\leftarrow)$\\\footnotesize{$p < 0.01$}} & 
            \makecell{$p: ns$} & 
            \makecell{$p: ns$} \\
        \cline{2-7}
        & \EFtt{1} &
            \makecell{\texttt{N/A}} & 
            \makecell{$p: ns$} & 
            \makecell{$p: ns$} & 
            \makecell{$\chi^2 = 25.9 ~(\leftarrow)$\\\footnotesize{$p < 0.001$}} & 
            \makecell{$\chi^2 = 29.0 ~(\leftarrow)$\\\footnotesize{$p < 0.001$}} \\
        \hline
    \end{tabular}
    \label{tab:swap_rates}
\end{table*}

\clearpage

\section{Factors Influencing Perceived Fairness}

\subsection{Feature Importance}
\label{app:feature_imp}


As discussed in \Cref{sec:h4_feature_explain}, we conduct an analysis of the influence of different features by fitting various machine learning models on a set of features that is determined by the amount of information available to participants in the corresponding treatment. \Cref{tab:factors_list} lists the features we consider for each case. The performance of each model is measured in terms of f1-score and accuracy, using leave-one-out cross-validation.

\begin{table*}[h]
\centering
\scriptsize
\caption{Factors considered in each of the five treatments.}
\begin{tabular}{lp{7cm}ccc}\toprule
Feature Name &Explanation &T1 &T2 \& T3 &T4 \& T5 \\\midrule
FracValue &The value the participant has for their bundle is divided by their total value for all items, i.e. $v_1(A_1)/v_1(M)$. &$\checkmark$ &$\checkmark$ &$\checkmark$ \\
NumItems &Number of items received by the participant, i.e. $|A_1|$. &$\checkmark$ &$\checkmark$ &$\checkmark$ \\
GetsHighest &Whether the participant receives the good they value the highest, i.e. $\mathds{1}{(j \in A_1)}$ where $j = \arg\max_{h \in M} v_{i,h}$. &$\checkmark$ &$\checkmark$ &$\checkmark$ \\
IsLarge &Whether the value received by the participant is greater than their `Large' threshold, i.e. $\mathds{1}{(v_1(A_1) \ge \frac{1}{2}}v_1(M))$. &$\checkmark$ &$\checkmark$ &$\checkmark$ \\
IsProp &Whether the value the participant has for their bundle is more than their proportional value, i.e. $\mathds{1}{(v_1(A_1) \ge v_1(M)/n)}$. &$\checkmark$ &$\checkmark$ &$\checkmark$ \\
IsProp1 &Whether the difference between the participant's value for their bundle and their proportional value is not more than the value they have for at least one good among the ones they don't receive, i.e. $\exists h \in M \backslash A_1$ such that $v_1(A_1 \cup \{h\}) \geq \frac{1}{n} v_1(M)$. &$\checkmark$ &$\checkmark$ &$\checkmark$ \\
IsMMS &Whether the participant's value is greater than their MMS threshold, i.e. $\mathds{1}{(v_1(A_1) \geq MMS_i^{(n)}(M))}$. &$\checkmark$ &$\checkmark$ &$\checkmark$ \\\midrule
MaxSelf &The maximum value the participant has for a bundle, i.e.$\max_{k \in N} v_1(A_k)$. & &$\checkmark$ &$\checkmark$ \\
MinSelf &The minimum value the participant has for a bundle, i.e. $\min_{k \in N} v_1(A_k)$. & &$\checkmark$ &$\checkmark$ \\
MeanSelf &The average of the participant's values for all bundles, i.e. $\frac{1}{n}*\sum_{k \in N} v_1(A_k)$. & &$\checkmark$ &$\checkmark$ \\
StdSelf &The standard deviation of the participant's values for all bundles, i.e.$\frac{1}{\sqrt{n}}*\sum_{k \in N} (v_1(A_k) - \mu)$, where $\mu$ corresponds to MeanSelf. & &$\checkmark$ &$\checkmark$ \\
BestSelfImprovement &The maximum gain the participant can experience by swapping their bundle (negative if not swapping is optimal), i.e. $\max_{k \in N \setminus \{i\}} v_1(A_k) - v_1(A_1)$. & &$\checkmark$ &$\checkmark$ \\\midrule
MaxAll &The maximum value an agent has for their respective bundle, i.e. $\max_{k \in N} v_k(A_k)$. & & &$\checkmark$ \\
MinAll &The minimum value an agent has for their respective bundle, i.e. $\min_{k \in N} v_k(A_k)$. & & &$\checkmark$ \\
MeanAll &The average of the value agents have for their respective bundles, i.e. $\frac{1}{n}*\sum_{k \in N} v_k(A_k)$. & & &$\checkmark$ \\
StdAll &The standard deviation of the value agents have for their respective bundles, i.e.$\frac{1}{\sqrt{n}}*\sum_{k \in N} (v_k(A_k) - \mu')$, where $\mu'$ corresponds to MeanAll. & & &$\checkmark$ \\
ParetoImprovementPossible &Whether there exists a swap such that the agent who the participant swaps with also benefits from it, i.e. $(\exists \{i,k\} \subset N : v_1(A_k) \ge v_1(A_1) \land v_k(A_1) \ge v_k(A_k))$. & & &$\checkmark$ \\
BestNetImprovement &Largest value of the improvement two agents (including the participant) can get from a swap, i.e. $\max_{\{i,k\} \subset N} v_1(A_k) - v_1(A_1) + v_k(A_1) - v_k(A_k)$. & & &$\checkmark$ \\
IsEF &Whether the participant values their bundle the most and no other agent envies the participant, i.e.$\mathds{1}{(\forall k \in N \setminus \{i\}, v_1(A_1) \ge v_1(A_k) \land v_k(A_k) \ge v_k(A_1))}$. Note this indicator considers only the bundle comparisons that are possible from the information available to the participants. & & &$\checkmark$ \\
\bottomrule
\end{tabular}
\label{tab:factors_list}
\end{table*}

Using logistic regression, we confirm that all features are indeed correlated with participants' perception of fairness. This means that features such as FracValue or IsMMS, where a larger value increases the chance of the participant considering the allocation fair, are associated with a positive weight in the LR model. Similarly, features such as BestSelfImprovement, where a larger value decreases the chance of the participant considering the allocation fair, are associated with a negative weight.

\begin{table*}[h]\centering
\scriptsize
\caption{Prediction accuracies and F1 scores for different ML models tested on participant decisions in each treatment. The `baseline' is a model that always predicts the majority class.}
\begin{tabular}{cccccccccccc}\toprule
&\multicolumn{5}{c}{\textbf{Accuracy}} & &\multicolumn{5}{c}{\textbf{F1 Score}} \\\cmidrule{2-6}\cmidrule{8-12}
\textbf{Treatment} &\textbf{LR} &\textbf{DT} &\textbf{RF} &\textbf{XGB} &\textbf{Baseline} & &\textbf{LR} &\textbf{DT} &\textbf{RF} &\textbf{XGB} &\textbf{Baseline} \\\midrule
\textbf{T1} &\textbf{0.84} &0.78 &0.79 &0.79 &0.69 & &\textbf{0.68} &0.59 &0.63 &0.62 &0 \\
\textbf{T2} &\textbf{0.83} &0.81 &0.8 &0.8 &0.81 & &\textbf{0.37} &0.29 &0.33 &0.29 &0 \\
\textbf{T3} &0.86 &0.85 &0.87 &\textbf{0.88} &0.66 & &0.9 &0.89 &0.9 &\textbf{0.91} &0 \\
\textbf{T4} &\textbf{0.66} &0.65 &0.64 &0.63 &0.65 & &0.31 &0.37 &\textbf{0.4} &0.39 &0 \\
\textbf{T5} &\textbf{0.77} &0.7 &0.72 &0.7 &0.72 & &\textbf{0.84} &0.8 &0.82 &0.8 &0 \\
\bottomrule
\end{tabular}
\label{tab:features_scores}
\end{table*}

We observe that these models (especially LR) yield significant improvements over a baseline model that always predicts the most common response from participants (fair or unfair). Interestingly, this improvement is much greater in the \textit{implicit} treatments (T2 and T4) as compared to the \textit{explicit} treatments (T3 and T5) (see the F1 scores in \Cref{tab:features_scores}), indicating that the predictability of decisions is impacted by the type of treatment. 

\Cref{tab:rf_features} and \Cref{tab:xgb_features} show the top-3 most important features as per Random Forest and XGBoost, respectively. As is the case with Decision Tree, \textit{FracValue} is the most influential feature when the fairness measure is implicit (with the exception of XGB in T4), while BestSelfImprovement is the most important feature when the fairness measure is explicit. 

\begin{table*}[h]\centering
\scriptsize
\caption{Top-3 most important features as per Random Forest, in terms of Gini index (indicated in brackets).}
\resizebox{\columnwidth}{!}{
\begin{tabular}{llllll}\toprule
\textbf{T1} &\textbf{T2} &\textbf{T3} &\textbf{T4} &\textbf{T5} \\\midrule
FracValue (0.58) &FracValue (0.27) &BestSelfImprovement (0.46) &FracValue (0.13) &BestSelfImprovement (0.22) \\
IsMMS (0.24) &BestSelfImprovement (0.12) &FracValue (0.18) &StdAll (0.09) &FracValue (0.11) \\
NumItems (0.08) &MinSelf (0.12) &IsProp (0.1) &MinAll (0.09) &IsEF (0.08) \\
\bottomrule
\end{tabular}}
\label{tab:rf_features}
\end{table*}

\begin{table*}[h]\centering
\scriptsize
\caption{Top-3 most important features as per XGBoost, in terms of Gini index (indicated in brackets).}
\resizebox{\columnwidth}{!}{
\begin{tabular}{llllll}\toprule
\textbf{T1} &\textbf{T2} &\textbf{T3} &\textbf{T4} &\textbf{T5} \\\midrule
FracValue (0.72) &FracValue (0.41) &BestSelfImprovement (0.81) &BestSelfImprovement (0.17) &BestSelfImprovement (0.47) \\
GetsHighest (0.17) &MeanSelf (0.13) &MeanSelf (0.06) &FracValue (0.14) &MaxAll (0.11) \\
NumItems (0.11) &StdSelf (0.13) &FracValue (0.03) &MinAll (0.13) &MinSelf (0.07) \\
\bottomrule
\end{tabular}}
\label{tab:xgb_features}
\end{table*}

\subsection{Motivating Swaps}\label{app:motivating_swaps}

As described in \Cref{sec:movtivating_swaps}, there is a significant decrease in the rate of {\em selfish swaps} when the participant is provided information about other agents' subjective values for goods in treatment T5 in comparison with treatment T3 where the participant has no information about other agents' valuations. We also observe a statistically significant increase in the fraction of {\em altruistic swaps} in T5 compared to T3. We determined the statistical significance of these observations in the relative differences in the fraction of instances where a selfish or altruistic swap was made (out of the total number of instances where selfish or altruistic swaps respectively that were possible) by comparing across both treatments as we summarize in \Cref{tab:swap_definitions}.

We also use the Fisher's exact test to compare the occurrence of \textit{Pareto improving} and \textit{utility-maximizing} swaps. \Cref{tab:pareto_utility} shows that differences in the fractions of either type of swap in T5 as compared to T3 are not statistically significant at  $p < 0.05$. Since participants do not have access to the valuations of other agents' in T3 (and hence cannot identify such swaps even if they are possible) but gain this information in T5, a possible explanation for this observation is that participants in our study did not factor welfare improvements into their decisions to swap even if they are made aware of the possibility of welfare improving swaps.

\begin{table*}[t]
\centering
\scriptsize
\caption{Rates of Pareto improving and utility maximizing swaps in treatments T3 and T5. The significance level for changes across both treatments is computed using Fisher's exact test and ``ns'' denotes $p\geq 0.05$.}
\begin{tabular}{cccccccccc}\toprule
&\multicolumn{3}{c}{\textbf{T3}} & &\multicolumn{3}{c}{\textbf{T5}} &  \\\cmidrule{2-4}\cmidrule{6-8}
\textbf{Type of swap} &\textbf{\makecell{Swaps\\made}} &\textbf{\makecell{Swaps\\possible}} &\textbf{\makecell{Percentage\\utilized}} & &\textbf{\makecell{Swaps\\made}} &\textbf{\makecell{Swaps\\possible}} &\textbf{\makecell{Percentage\\utilized}} &\textbf{\makecell{Significant\\difference}} \\\midrule
\textbf{Pareto improving} &30 &43 &69.76 & &19 &35 &54.28 &$p : ns$ \\
\textbf{Utility-maximizing} &66 &96 &68.75 & &52 &89 &58.42 &$p : ns$\\
\bottomrule
\end{tabular}
\label{tab:pareto_utility}
\end{table*}

\subsection{Relationship with Cognitive Effort}
\label{apx:cog_effort_charts}

\Cref{tab:time_to_answer} and \Cref{tab:feedback} show that we do not find sufficient evidence establish to establish whether the treatment participants are subjected to has an effect on either response times or reported difficulty respectively. As described in \Cref{sec:cognitive}, we do not find enough evidence to suggest that \textit{bundle information} information has any effect on either of these metrics. We draw this conclusion using Welch's $t$-test, where an increase in bundle information (from T1 to T2 or T3) does not result in a significant change ($p \ge 0.05$) in participants' response times (in seconds) and reported difficulty levels (on a 5-point Likert scale). Using the same test, we observe no influence of the \textit{fairness measure} either (by comparing T2 with T3, and T4 with T5), on either metric. 

\begin{table*}[h]
    \scriptsize
    \centering
    \caption{ $p$-values of the $t$ statistic for testing equal means of participant response times per scenario using Welch's $t$-test -- for different pairs of treatments, and adjusting for different variables. Here, ``ns'' denotes $p\geq 0.05$.
    }
    \begin{tabular}{|c|c||c|c|c|c|c|c|c|c|c|c|}\hline
        \multirow{2}{*}{\bf Variable} & \multirow{2}{*}{\bf Value} & \multicolumn{6}{c|}{\bf Pairs of Treatments} \\ \cline{3-8}
         & & T1 vs. T2 & T1 vs. T4 & T2 vs. T3 & T2 vs. T4 & T3 vs. T5 & T4 vs. T5 \\ \hline \hline
        \multicolumn{2}{|c||}{All scenarios} & 
            \makecell{$p: ns$} & 
            \makecell{$p: ns$} & 
            \makecell{$p: ns$} & 
            \makecell{$p < 0.05$} & 
            \makecell{$p < 0.01$} & 
            \makecell{$p: ns$} \\[0.5em]
        \hline
        \multirow{3}{*}{\makecell{Allocation\\ Type}} 
        & \Proptt{} &
            \makecell{$p: ns$} & 
            \makecell{$p: ns$} & 
            \makecell{$p: ns$} & 
            \makecell{$p: ns$} & 
            \makecell{$p: ns$} & 
            \makecell{$p: ns$} \\[0.5em]
        \cline{2-8}
        & \Proptt{1} &
            \makecell{$p: ns$} & 
            \makecell{$p: ns$} & 
            \makecell{$p: ns$} & 
            \makecell{$p: ns$} & 
            \makecell{$p < 0.05$} & 
            \makecell{$p: ns$} \\[0.5em]
        \cline{2-8}
        & \MMStt{} &
            \makecell{$p: ns$} & 
            \makecell{$p: ns$} & 
            \makecell{$p: ns$} & 
            \makecell{$p: ns$} & \
            \makecell{$p: ns$} & 
            \makecell{$p: ns$} \\[0.5em]
        \hline
    \end{tabular}
    \label{tab:time_to_answer}
\end{table*}

\begin{table*}[h] 
    \scriptsize
    \centering
\caption{ $p$-values of the $t$ statistic for testing equal means of participant reported difficulty using Welch's $t$-test -- for different pairs of treatments where ``ns'' denotes $p\geq 0.05$.
}
    \begin{tabular}{|c|c|c|c|c|}
        \hline
        \textbf{Pairs of Treatments} & T2 & T3 & T4 & T5 \\
         \hline
        \makecell{T1} & \makecell{{$p: ns$}} & \makecell{{$p: ns$}} & \makecell{{$p < 0.001$}} & \makecell{{$p < 0.001$}}  \\ 
        \makecell{T2} & \cellcolor{lightgray} & \makecell{{$p: ns$}} & \makecell{{$p < 0.01$}} & \makecell{{$p < 0.05$}}  \\ 
        \makecell{T3} & \cellcolor{lightgray} & \cellcolor{lightgray} & \makecell{{$p < 0.05$}} & \makecell{{$p:ns$}}  \\ 
        \makecell{T4} & \cellcolor{lightgray} & \cellcolor{lightgray} & \cellcolor{lightgray} & \makecell{{$p:ns$}}  \\ 
         \hline
    \end{tabular}
    \label{tab:feedback}
\end{table*}

However, an increase in \textit{value information} does lead to a significant increase in response time at a significance level of $p < 0.05$ between T2 and T4, and at $p < 0.01$ between T3 and T5 using Welch's t-test. While there is a significant increase in reported difficulty between T2 and T4 (at $p < 0.01$), there is no significant change between T3 and T5. Considering all of these observations, we do not reject the null hypothesis (\Cref{sec:cognitive}) that \textit{cognitive effort does not depend on treatment}.

\end{document}